\documentclass[conference]{IEEEtran}
\usepackage{url}
\usepackage[utf8]{inputenc}
\usepackage{xcolor}
\usepackage{amsmath}

\usepackage{multirow}
\usepackage{amssymb}
\usepackage{cite}
\usepackage{enumitem}
\usepackage{bm}
\usepackage{graphicx}
\usepackage{amsthm}

\usepackage{gensymb}

\usepackage[acronyms,nonumberlist,nopostdot,nomain,nogroupskip]{glossaries}
\usepackage{tablefootnote}
\usepackage{booktabs}
\usepackage{tabularx}
\usepackage{epsfig}
\usepackage{tikz}
\usepackage{pgfplots}
\pgfplotsset{compat=newest} 
\pgfplotsset{plot coordinates/math parser=false}

\usetikzlibrary{plotmarks,patterns,patterns.meta,decorations.pathreplacing,backgrounds,calc,arrows,arrows.meta,spy,matrix}
\usepgfplotslibrary{patchplots,groupplots}
\usepackage{tikzscale}
\usepackage{siunitx}
\usepackage{tablefootnote}

\usepackage{multirow}
\usepackage{algorithm2e, setspace}
\SetAlCapNameFnt{\footnotesize}
\SetAlCapFnt{\footnotesize}
\RestyleAlgo{ruled}
\SetKwComment{Comment}{/* }{ */}
\usepackage{algpseudocode}

\usepackage[font=scriptsize]{subcaption}
\usepackage[font=footnotesize]{caption}

\usepackage{mathtools}

\usepackage{dblfloatfix}    
\usepackage{colortbl}
\usepackage{flushend}
\usepackage{footnote}

\usepackage{lipsum,afterpage,refcount}

\newacronym{3gpp}{3GPP}{3rd Generation Partnership Project}
\newacronym{adc}{ADC}{Analog to Digital Converter}
\newacronym{5g}{5G}{5th generation}
\newacronym{6g}{6G}{6th generation}
\newacronym{ai}{AI}{Artificial Intelligence}
\newacronym{aimd}{AIMD}{Additive Increase Multiplicative Decrease}
\newacronym{am}{AM}{Acknowledged Mode}
\newacronym{amc}{AMC}{Adaptive Modulation and Coding}
\newacronym{aqm}{AQM}{Active Queue Management}
\newacronym{awgn}{AGWN}{Additive White Gaussian Noise}
\newacronym{balia}{BALIA}{Balanced Link Adaptation}
\newacronym{bdp}{BDP}{Bandwidth-Delay Product}
\newacronym{bf}{BF}{beamforming}
\newacronym{cc}{CC}{Congestion Control}
\newacronym{cdf}{CDF}{Cumulative Distribution Function}
\newacronym{cn}{CN}{Core Network}
\newacronym{cqi}{CQI}{Channel Quality Information}
\newacronym{cp}{CP}{Control Plane}
\newacronym{csirs}{CSI-RS}{Channel State Information - Reference Signal}
\newacronym{dc}{DC}{Dual Connectivity}
\newacronym{rb}{RB}{Resource Block}
\newacronym{dce}{DCE}{Direct Code Execution}
\newacronym{dci}{DCI}{Downlink Control Information}
\newacronym{udp}{UDP}{User Datagram Protocol}
\newacronym{dl}{DL}{downlink}
\newacronym{fcfs}{FCFS}{first-come-first-served}
\newacronym{dmr}{DMR}{Deadline Miss Ratio}
\newacronym{fspl}{FSPL}{free-space path loss}
\newacronym{dmrs}{DMRS}{DeModulation Reference Signal}
\newacronym{e2e}{E2E}{End-to-End}
\newacronym{ppp}{PPP}{Poission Point Process}
\newacronym{aoi}{AoI}{Area of Interest}
\newacronym{cpu}{CPU}{Central Processing Unit}
 \newacronym{gpu}{GPU}{Graphics Processing Unit}
 \newacronym{tpu}{TPU}{Tensor Processing Unit}
\newacronym{si}{SI}{Study Item}
\newacronym{ecn}{ECN}{Explicit Congestion Notification}
\newacronym{edf}{EDF}{Earliest Deadline First}
\newacronym{enb}{eNB}{eNodeB}
\newacronym{epc}{EPC}{Evolved Packet Core}
\newacronym{es}{ES}{Edge Server}
\newacronym{cav}{CAV}{Connected and Autonomous Vehicle}
\newacronym{fdma}{FDMA}{Frequency Division Multiple Access}
\newacronym{fdd}{FDD}{Frequency Division Duplexing}
\newacronym{upa}{UPA}{Uniform Planar Array}
\newacronym{car}{CAR}{Circular Aperture Reflector }
\newacronym[firstplural=Radio Access Technologies (RATs)]{rat}{RAT}{Radio Access Technology}
\newacronym[firstplural=Radio Access Technology (RTs)]{rt}{RT}{Radio Technology}
\newacronym{fs}{FS}{Fast Switching}
\newacronym{isd}{ISD}{inter-site distance}
\newacronym{ftp}{FTP}{File Transfer Protocol}
\newacronym{gnb}{gNB}{Next Generation Node Base}
\newacronym{harq}{HARQ}{Hybrid Automatic Repeat reQuest}
\newacronym{hetnet}{HetNet}{Heterogeneous Network}
\newacronym{hh}{HH}{Hard Handover}
\newacronym{hol}{HOL}{Head-of-Line}
\newacronym{ia}{IA}{Initial Access}
\newacronym{imt}{IMT}{International Mobile Telecommunication}
\newacronym{iot}{IoT}{Internet of Things}
\newacronym{los}{LOS}{Line of Sight}
\newacronym{lte}{LTE}{Long Term Evolution}
\newacronym{m2m}{M2M}{Machine to Machine}
\newacronym{mac}{MAC}{Medium Access Control}
\newacronym{mc}{MC}{Multi-Connectivity}
\newacronym{mcs}{MCS}{Modulation and Coding Scheme}
\newacronym{mec}{MEC}{Mobile Edge Cloud}
\newacronym{mi}{MI}{Mutual Information}
\newacronym{mimo}{MIMO}{Multiple Input Multiple Output}
\newacronym{mmwave}{mmWave}{millimeter wave}
\newacronym{mptcp}{MPTCP}{Multipath TCP}
\newacronym{mr}{MR}{Maximum Rate}
\newacronym{mss}{MSS}{Maximum Segment Size}
\newacronym{mtd}{MTD}{Machine-Type Device}
\newacronym{mtu}{MTU}{Maximum Transmission Unit}
\newacronym{nfv}{NFV}{Network Function Virtualization}
\newacronym{vnf}{VNF}{Virtualization Network Function}
\newacronym{gv}{GV}{ground vehicle}
\newacronym{vec}{VEC}{Vehicular Edge Computing}
\newacronym{sdn}{SDN}{Software Defined Networking}
\newacronym{nlos}{NLOS}{Non Line of Sight}
\newacronym{nlosb}{NLOSb}{Building Non Line of Sight}
\newacronym{nlosv}{NLOSv}{Vehicle Non Line of Sight}
\newacronym{nr}{NR}{New Radio}
\newacronym{ofdm}{OFDM}{Orthogonal Frequency Division Multiplexing}
\newacronym{pdcch}{PDCCH}{Physical Downlonk Control Channel}
\newacronym{pdcp}{PDCP}{Packet Data Convergence Protocol}
\newacronym{pdsch}{PDSCH}{Physical Downlink Shared Channel}
\newacronym{pdu}{PDU}{Packet Data Unit}
\newacronym{pf}{PF}{Proportional Fair}
\newacronym{pgw}{PGW}{Packet Gateway}
\newacronym{phy}{PHY}{Physical}
\newacronym{pbch}{PBCH}{Physical Broadcast Channel}
\newacronym[plural=\gls{mme}s,firstplural=Mobility Management Entities (MMEs)]{mme}{MME}{Mobility Management Entity}
\newacronym{prb}{PRB}{Physical Resource Block}
\newacronym{pss}{PSS}{Primary Synchronization Signal}
\newacronym{pucch}{PUCCH}{Physical Uplink Control Channel}
\newacronym{pusch}{PUSCH}{Physical Uplink Shared Channel}
\newacronym{rach}{RACH}{Random Access Channel}
\newacronym{ran}{RAN}{Radio Access Network}
\newacronym{red}{RED}{Random Early Detection}
\newacronym{rf}{RF}{Radio Frequency}
\newacronym{rlc}{RLC}{Radio Link Control}
\newacronym{rlf}{RLF}{Radio Link Failure}
\newacronym{rrc}{RRC}{Radio Resource Control}
\newacronym{rrm}{RRM}{Radio Resource Management}
\newacronym{rr}{RR}{Round Robin}
\newacronym{rs}{RS}{Remote Server}
\newacronym{rsrp}{RSRP}{Reference Signal Received Power}
\newacronym{rss}{RSS}{Received Signal Strength}
\newacronym{rtt}{RTT}{Round Trip Time}
\newacronym{rw}{RW}{Receive Window}
\newacronym{rx}{RX}{Receiver}
\newacronym{sa}{SA}{standalone}
\newacronym{sack}{SACK}{Selective Acknowledgment}
\newacronym{sap}{SAP}{Service Access Point}
\newacronym{sch}{SCH}{Secondary Cell Handover}
\newacronym{scoot}{SCOOT}{Split Cycle Offset Optimization Technique}
\newacronym{sdma}{SDMA}{Spatial Division Multiple Access}
\newacronym{sinr}{SINR}{Signal to Interference plus Noise Ratio}
\newacronym{sm}{SM}{Saturation Mode}
\newacronym{snr}{SNR}{Signal-to-Noise Ratio}
\newacronym{son}{SON}{Self-Organizing Network}
\newacronym{ss}{SS}{Synchronization Signal}
\newacronym{srs}{SRS}{Sounding Reference Signal}
\newacronym{sss}{SSS}{Secondary Synchronization Signal}
\newacronym{tb}{TB}{Transport Block}
\newacronym{tcp}{TCP}{Transmission Control Protocol}
\newacronym{tdd}{TDD}{Time Division Duplexing}
\newacronym{tdma}{TDMA}{Time Division Multiple Access}
\newacronym{tfl}{TfL}{Transport for London}
\newacronym{tm}{TM}{Transparent Mode}
\newacronym{prr}{PRR}{Packet Reception Ratio}
\newacronym{trp}{TRP}{Transmitter Receiver Pair}
\newacronym{tti}{TTI}{Transmission Time Interval}
\newacronym{ttt}{TTT}{Time-to-Trigger}
\newacronym{tx}{TX}{Transmitter}
\newacronym{ue}{UE}{User Equipment}
\newacronym{ul}{UL}{uplink}
\newacronym{uml}{UML}{Unified Modeling Language}
\newacronym{um}{UM}{Unacknowledged Mode}
\newacronym{utc}{UTC}{Urban Traffic Control}
\newacronym{vm}{VM}{Virtual Machine}
\newacronym{rsrq}{RSRQ}{Reference Signal Received Quality}
\newacronym{rssi}{RSSI}{Received Signal Strength Indicator}
\newacronym{crs}{CRS}{Cell Reference Signal}
\newacronym{v2v}{V2V}{Vehicle-to-Vehicle}
\newacronym{v2i}{V2I}{Vehicle-to-Infrastructure}
\newacronym{v2n}{V2N}{Vehicle-to-Network}
\newacronym{v2x}{V2X}{Vehicle-to-Everything}
\newacronym{vn}{VN}{Vehicular Node}
\newacronym{dsrc}{DSRC}{Dedicated Short Range Communication}
\newacronym{ci}{CI}{context information}
\newacronym{voi}{VoI}{value of information}
\newacronym{gps}{GPS}{Global Positioning System}
\newacronym{qos}{QoS}{Quality of Service}
\newacronym{qoe}{QoE}{Quality of Experience}
\newacronym{ml}{ML}{Machine Learning}
\newacronym{ahp}{AHP}{Analytic Hierarchy Process}
\newacronym{lidar}{LIDAR}{Light Detection and Ranging}
\newacronym{sumo}{SUMO}{Simulation of Urban MObility}
\newacronym{wave}{WAVE}{Wireless Access in Vehicular Environment}
\newacronym{c-its}{C-ITS}{Connected Intelligent Transportation System}
\newacronym{dash}{DASH}{Dynamic Adaptive Streaming over HTTP}
\newacronym{http}{HTTP}{HyperText Transfer Protocol}
\newacronym{nt}{NT}{Non-Terrestrial}
\newacronym{ntc}{NTC}{non-terrestrial communication}
\newacronym{ntn}{NTN}{Non-Terrestrial Network}
\newacronym{haps}{HAPS}{High Altitude Platform Station}
\newacronym{hap}{HAP}{High Altitude Platform}
\newacronym{leo}{LEO}{Low Earth Orbit}
\newacronym{meo}{MEO}{Medium Earth Orbit}
\newacronym{geo}{GEO}{Geostationary Earth Orbit}
\newacronym{uav}{UAV}{Unmanned Aerial Vehicle}
\newacronym{nsat}{nSAT}{Nanosatellite}
\newacronym{ehf}{EHF}{extremely high-frequency}
\newacronym{ioe}{IoE}{Internet of Everyone}
\newacronym{gan}{GaN}{Gallium Nitride}

\usepackage{tikz}
\usepackage{pgfplots}
\usepackage{tikzscale}
\usepackage{tikz-qtree}

\pgfplotsset{compat=newest}
\pgfplotsset{plot coordinates/math parser=false}
\pgfplotsset{every axis/.append style={
                    label style={font=\scriptsize},
                    tick label style={font=\scriptsize},
                    legend style={font=\scriptsize}
                    }}
\usetikzlibrary{plotmarks,shapes,patterns,decorations.pathreplacing,backgrounds,calc,arrows,arrows.meta,spy,matrix,shadows,trees,positioning,fit}
\usepgfplotslibrary{patchplots,groupplots}

\tikzstyle{startstop} = [rectangle, rounded corners, minimum width=2cm, minimum height=0.5cm,text centered, draw=black]
\tikzstyle{io} = [trapezium, trapezium left angle=70, trapezium right angle=110, minimum width=3cm, minimum height=1cm, text centered, draw=black]
\tikzstyle{process} = [rectangle, minimum width=2cm, minimum height=0.5cm, text centered, draw=black, alignb=center]
\tikzstyle{decision} = [ellipse, minimum width=2cm, minimum height=1cm, text centered, draw=black]
\tikzstyle{arrow} = [thick,<->,>=stealth]
\tikzstyle{line} = [thick,>=stealth]
\tikzstyle{darrow} = [thick,<->,>=stealth,dashed]
\tikzstyle{sarrow} = [thick,->,>=stealth]
\tikzstyle{larrow} = [line width=0.1mm,dashdotted,->,>=stealth]

\pgfkeys{/pgf/number format/.cd,1000 sep={\,}} 

\makeatletter
\def\grd@save@target#1{%
  \def\grd@target{#1}}
\def\grd@save@start#1{%
  \def\grd@start{#1}}
\tikzset{
  grid with coordinates/.style={
    to path={%
      \pgfextra{%
        \edef\grd@@target{(\tikztotarget)}%
        \tikz@scan@one@point\grd@save@target\grd@@target\relax
        \edef\grd@@start{(\tikztostart)}%
        \tikz@scan@one@point\grd@save@start\grd@@start\relax
        \draw[minor help lines] (\tikztostart) grid (\tikztotarget);
        \draw[major help lines] (\tikztostart) grid (\tikztotarget);
        \grd@start
        \pgfmathsetmacro{\grd@xa}{\the\pgf@x/1cm}
        \pgfmathsetmacro{\grd@ya}{\the\pgf@y/1cm}
        \grd@target
        \pgfmathsetmacro{\grd@xb}{\the\pgf@x/1cm}
        \pgfmathsetmacro{\grd@yb}{\the\pgf@y/1cm}
        \pgfmathsetmacro{\grd@xc}{\grd@xa + \pgfkeysvalueof{/tikz/grid with coordinates/major step x}}
        \pgfmathsetmacro{\grd@yc}{\grd@ya + \pgfkeysvalueof{/tikz/grid with coordinates/major step y}}
        \foreach \x in {\grd@xa,\grd@xc,...,\grd@xb}
        \node[anchor=north] at (\x,\grd@ya) {\pgfmathprintnumber{\x}};
        \foreach \y in {\grd@ya,\grd@yc,...,\grd@yb}
        \node[anchor=east] at (\grd@xa,\y) {\pgfmathprintnumber{\y}};
      }
    }
  },
  minor help lines/.style={
    help lines,
    gray,
    line cap =round,
    xstep=\pgfkeysvalueof{/tikz/grid with coordinates/minor step x},
    ystep=\pgfkeysvalueof{/tikz/grid with coordinates/minor step y}
  },
  major help lines/.style={
    help lines,
    line cap =round,
    line width=\pgfkeysvalueof{/tikz/grid with coordinates/major line width},
    xstep=\pgfkeysvalueof{/tikz/grid with coordinates/major step x},
    ystep=\pgfkeysvalueof{/tikz/grid with coordinates/major step y}
  },
  grid with coordinates/.cd,
  minor step x/.initial=.5,
  minor step y/.initial=.2,
  major step x/.initial=1,
  major step y/.initial=1,
  major line width/.initial=1pt,
}
\makeatother

\newlength\fheight
\newlength\fwidth

\definecolor{steelblue}{RGB}{176,196,222}

\usepackage{hyperref}
\usepackage[capitalize]{cleveref}
\crefname{section}{Sec.}{Secs.}
\usetikzlibrary{decorations}

\makeglossaries
\begin{document}
\bstctlcite{IEEEexample:BSTcontrol}

\title{On the Energy Consumption of UAV Edge Computing in Non-Terrestrial Networks}

    \author{\IEEEauthorblockN{Alessandro Traspadini$^{\circ }$, Marco Giordani$^{\circ }$, Giovanni Giambene$^{\star }$, Tomaso De Cola$^{\dag }$, Michele Zorzi$^{\circ }$\medskip}
\IEEEauthorblockA{
$^{\circ}$University of Padova, Italy. Email: \texttt{\{traspadini,giordani,zorzi\}@dei.unipd.it}\\
$^{\star}$University of Siena. Email: \texttt{\{giovanni.giambene\}@unisi.it}\\
$^{\dag}$DLR -- Institute of Communications and Navigation, Germany. Email: \texttt{\{tomaso.decola\}@dlr.de}
}}



\maketitle

\begin{abstract}
During the last few years, \glspl{uav} equipped with sensors and cameras have emerged as a cutting-edge technology to provide services such as surveillance, infrastructure inspections, and target acquisition.
However, this approach requires \glspl{uav} to process data onboard, mainly for person/object detection and recognition, which may pose significant energy constraints as \glspl{uav} are battery-powered. 
A possible solution can be the support of \glspl{ntn} for edge computing. 
In particular, \glspl{uav} can partially offload data (e.g., video acquisitions from onboard sensors) to more powerful upstream \glspl{hap} or satellites acting as edge computing servers to increase the battery autonomy compared to local processing, even though at the expense of some data transmission delays.
Accordingly, in this study we model the energy consumption of \glspl{uav}, \glspl{hap}, and satellites considering the energy for data processing, offloading, and hovering. Then, we investigate whether data offloading can improve the system performance. 
Simulations demonstrate that edge computing can improve both UAV autonomy and end-to-end delay compared to onboard processing in many configurations. 
\end{abstract}


\begin{tikzpicture}[remember picture,overlay]
\node[anchor=north,yshift=-10pt] at (current page.north) {\parbox{\dimexpr\textwidth-\fboxsep-\fboxrule\relax}{
\centering\footnotesize This paper has been accepted for publication at the 57th Asilomar Conference on Signals, Systems, and Computers. \textcopyright 2023 IEEE.\\
Please cite it as: A. Traspadini, M. Giordani, G. Giambene, T. De Cola, and M. Zorzi, "On the Energy Consumption of UAV Edge Computing in Non-Terrestrial Networks,"  57th Asilomar Conference on Signals, Systems, and Computers, 2023.}};
\end{tikzpicture}

\glsresetall

\section{Introduction}
\label{sec:intro}
In the context of \gls{6g} wireless networks~\cite{giordani2020toward}, \gls{ntn} is an emerging paradigm that consists of deploying aerial/space nodes such as \glspl{uav}, \glspl{hap} and satellites to improve the coverage and capacity of ground users when terrestrial infrastructure is not deployed, unavailable, or overloaded~\cite{giordani2021non}.
In particular, given their low cost and their versatility, \glspl{uav} can support applications such as surveillance, traffic control, disaster management, environmental monitoring, and high-precision agriculture~\cite{8761851,Alsalam17Autonomous}.
To enable these services, \glspl{uav} need to store and process data generated from local sensors (e.g., video acquisitions from video cameras), mainly to detect and recognize critical objects and/or potential dangers on the ground~\cite{Zhang20Beyond,bordin2022autonomous}.
However, data processing often involves running computationally demanding deep learning and computer vision algorithms onboard the \gls{uav}~\cite{Zhang20Beyond}, which may not be compatible with the limited computational capacity and battery lifetime of these platforms. In fact, \glspl{uav} consume significant energy for hovering and propulsion, and data processing will further reduce their autonomy.

To solve this issue, UAVs can delegate the burden of data processing to more powerful computing servers located at the edge of the network, a paradigm referred to as (mobile) edge computing~\cite{Khan19Edge}. In this way, rotary-wing \glspl{uav} consume less power, and the processing can be performed even if the \gls{uav} has no processing units~\cite{Kim19Optimal}.
In these regards, the best choice would be to deploy the edge servers as close to the end users as possible to reduce the delays for data offloading and for the processed output to be returned. 
Edge servers can also be located on \gls{ntn} nodes, for example on \glspl{hap} or \gls{leo} satellites~\cite{traspadini2022uavhapassisted}, so as to provide large and continuous geographical coverage even in rural/remote areas~\cite{Chaoub20216g} or in the absence of pre-existing terrestrial infrastructures~\cite{Liu22Mec}.
In our previous works, we studied the potential of NTN-assisted edge computing applied to vehicular~\cite{Traspadini23Real} and \gls{iot}~\cite{wang2022performance} networks. Specifically, we optimized the offloading rate to maximize the probability of real-time service given delay and computational capacity constraints.
The integration of \gls{leo} satellites with \glspl{uav} has also been studied in~\cite{Chen21Energy}, where the authors optimized the offloading rate to enable faster data processing.
However, most of the literature tends to consider data generated on the ground and relayed to (or processed by) the \gls{uav}, and measure the performance of edge computing only in terms of delay, with limited considerations related to energy consumption. 

To fill these gaps, in this paper we analyze a scenario where a swarm of \glspl{uav} collects sensory data that needs to be processed by an object detection algorithm. The processing can be performed locally onboard the \glspl{uav}, or data can be offloaded to an NTN edge computing server (either a \gls{hap} or a \gls{leo} satellite). Then, we develop an energy model to characterize the energy consumption of \glspl{uav}, \glspl{hap}, and LEO satellites for data processing and offloading, as well as for hovering and/or movement. Finally, we model the system as a set of D/M/1 queues, and investigate the impact of data offloading on the UAV battery autonomy and the end-to-end delay.
We demonstrate that both HAP- and LEO-assisted edge computing can improve the autonomy and delay compared to onboard processing in many configurations.

The rest of the paper is organized as follows. 
In Sec.~\ref{sec:system_model} we formulate our research problem and present our channel and delay models. In Sec.~\ref{sec:energy_model} we describe our energy model. In Sec.~\ref{sec:performance_evaluation} we provide numerical results. In Sec.~\ref{sec:conclusions_and_future_works} we conclude the paper with suggestions for future work.

\begin{figure}[t!]
\centering 
\includegraphics[width=0.9\columnwidth]{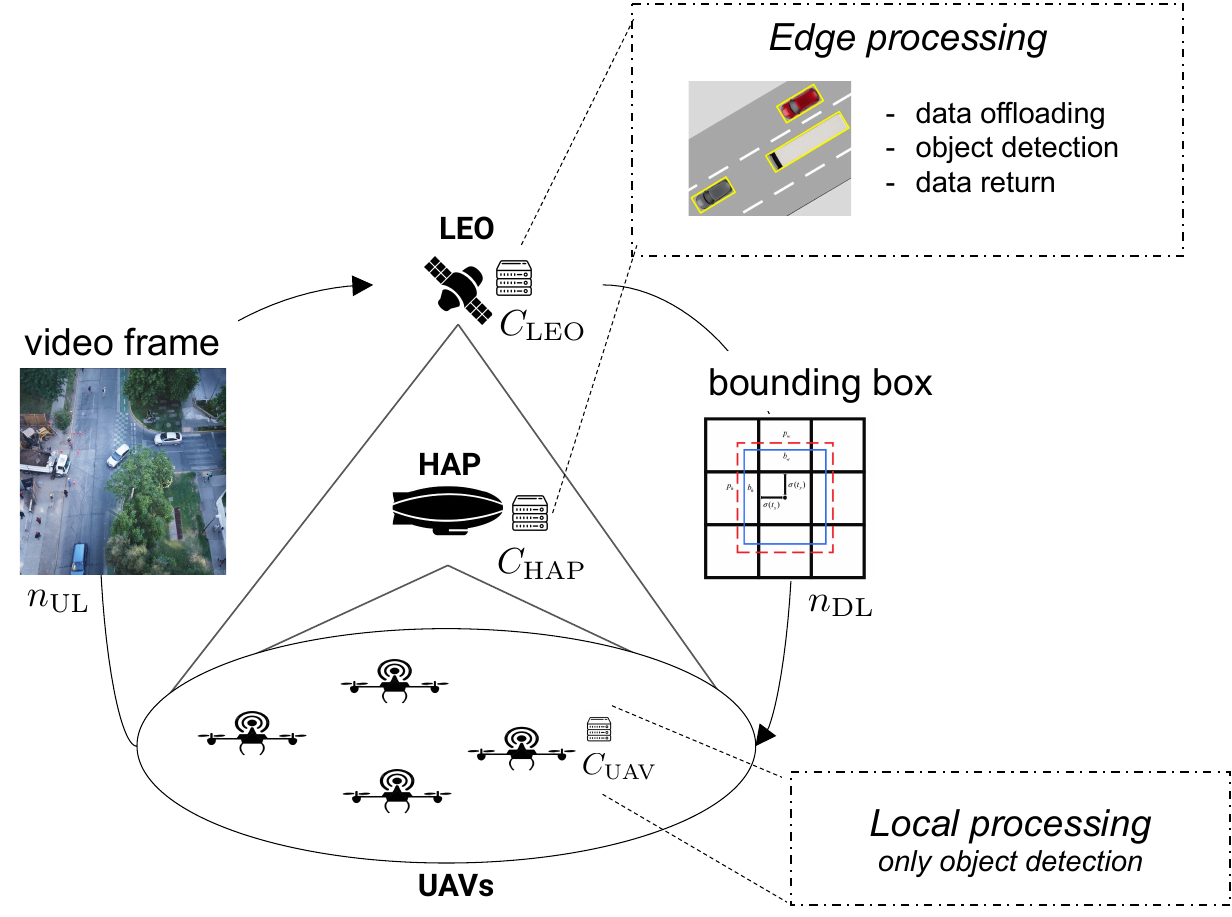}
\caption{The edge computing scenario. UAVs are equipped with video cameras recording video frames of size $s_{\rm UL}$ at rate $r$, that may or may not be offloaded to a HAP or a LEO satellite for processing. The processed output is of size $s_{\rm DL}$. The computational capacity is $C_{i}$, $i\in{\{\text{UAV, HAP, LEO}\}}$.\vspace{-0.5cm}}
\label{fig:uav_scenario}
\end{figure}

\section{System Model}
\label{sec:system_model}
In this section we present our research problem (Sec.~\ref{sub:formulation}), and our channel (Sec.~\ref{sub:channel}) and delay (Sec.~\ref{sub:queue}) models.

\subsection{Problem Formulation}
\label{sub:formulation}
We consider a set of $n$ \glspl{uav} equipped with video cameras that record video frames of size $s_{\rm UL}$ at an average frame rate $r$.
Eventually, video frames have to be processed by an object detection algorithm to detect and recognize critical entities in the surrounding environment, which results in a constant computational load $C_l$, expressed in Giga Floating Point Operations (GFLOP), which counts the number of floating point operations required to complete a processing task. 
The processed output (in this study, the bounding boxes of the detected entities) is returned in a packet of size $s_{\rm DL}\ll s_{\rm UL}$.
As shown in~\cref{fig:uav_scenario}, the area is monitored by an upstream \gls{hap} or \gls{leo} satellite,\footnote{We assume that the time to upload a single video frame from the UAV to the satellite is less than the visibility period of the satellite, i.e., a few minutes.} which may also act as edge server for the UAVs. Therefore, data processing can be performed either onboard the \glspl{uav} (i.e., local processing), or at the \gls{hap} or the \gls{leo} satellite (i.e., edge computing), or as a combination of the two.

The computational capacity, expressed in GFLOP/s and defined in terms of the number of operations that can be processed in one second, is $C_{i}$, $i\in{\{\text{UAV, HAP, LEO}\}}$, with $C_{\rm UAV}<C_{\rm HAP}<C_{\rm LEO}$ due to energy and space constraints. As such, edge computing is faster, at the expense of a non-negligible communication delay for offloading data from the UAVs to the edge server, and for the processed output to be returned. 
The offloading factor $\eta$ indicates the probability of a frame being offloaded to the \gls{hap} or the \gls{leo} satellite  for edge computing.
Therefore, the average arrival rate for data processing at the edge server is $\eta n r$, while for local processing at each \gls{uav} it is $(1-\eta)r$. 
In this paper, we will evaluate the energy consumption and  end-to-end delay for local processing vs. edge computing as a function of $\eta$.



\subsection{Channel Model}
\label{sub:channel}
We assume that the \glspl{uav} and the \gls{hap} or the \gls{leo} satellite communicate in the \gls{mmwave} bands, and the channel is modeled based on the \gls{3gpp} specifications for NTN~\cite{38821,giordani2020satellite}.
Assuming an interference-free scenario with \glspl{uav} operating in orthogonal frequency bands, the \gls{snr} between the UAV and receiver $i\in\{\text{HAP, LEO}\}$ is
\begin{equation}
\label{eq:snr}
\gamma_{i} = \frac{\text{EIRP} \cdot ({G}/{T})_i}{\text{PL}_{i} \cdot K_b \cdot B}, \: i\in\{\text{HAP, LEO}\}
\end{equation}
where $\text{EIRP}$ is the effective isotropic radiated power, $({G}/{T})_i$ is the receiver antenna-gain-to-noise-temperature, $\text{PL}_{i}$ is the path loss, $K_b$ is the Boltzmann constant, and $B$ is the bandwidth.
The path loss depends on the carrier frequency and the distance between transmitter and receiver, 
and accounts for additional atmospheric losses such as scintillation as described in~\cite{giordani2020satellite}.

The ergodic (Shannon) capacity $R_{i}$ can be derived~as
\begin{equation}
\label{eq:capacity}
R_{i}=B\log_2(1+\gamma_{i}), \: i\in\{\text{HAP, LEO}\}
\end{equation}


\subsection{End-to-End Delay Model}
\label{sub:queue}
The system is modeled as a set of D/M/1 queues.
At each UAV, the arrival rate is a function of the offloading factor $\eta$ and is equal to $\lambda_{\rm UAV}=(1-\eta)r$, while the service rate is $\mu_{\rm UAV}= C_{\rm UAV}/C_l$.
At the HAP/LEO edge server, the arrival rate is equal to $\lambda_{i} = \eta r n$ frames/s, while the service rate is $\mu_{i}= C_{i}/C_l$ frames/s, $i\in\{\text{HAP, LEO}\}$. 
According to~\cite{Jan66Queue}, the average system time (the sum of waiting and service times) for a D/M/1 queue with arrival rate $\lambda$ and service rate $\mu$ is
\begin{equation}
t_s (\lambda, \mu) = \frac{1}{\mu(1-\delta)},
\end{equation}
where $\delta$ is the root of the equation $\delta = e^{-\mu / \lambda(1-\delta)}$, to be solved numerically.\footnote{We calculated the roots of $\delta$ in the range of interest in terms of $\mu$ and $\lambda$ using Newton's method, and we obtained	$\delta\simeq4.2e^{-(3\mu)/(2\lambda)}$.}
We distinguish between local processing and edge computing.

\begin{itemize}
\item \emph{Local processing.}
The end-to-end delay is only due to data processing, and can be computed as 
\begin{equation}
    \Bar{t}_{\rm UAV} = t_s (\lambda_{\rm UAV}, \mu_{\rm UAV}).
\end{equation}

\item \emph{Edge computing.}
The end-to-end delay is the sum of the delay to offload video frames from the UAV to the edge server, the delay for data processing, and the delay for the processed output to be returned to the UAV, i.e.,
\begin{equation}
    \Bar{t}_i = 2\tau^{i}_p + t_{\rm UL}^i + t_{\rm DL}^i + t_s (\lambda_{i}, \mu_{i}), \: i\in\{\text{HAP, LEO}\}.
    \label{eq:edge-processing}
\end{equation}
In~\cref{eq:edge-processing}, $\tau^{i}_p=d_i/c$ is the propagation delay, where $d_i$ is the distance between the \gls{uav} and the edge server $i\in\{\text{HAP, LEO}\}$, and $c$ is the speed of light. Moreover, $t_{\rm UL}^i$ and $t_{\rm DL}^i$ are the transmission delays to send frames to the edge server and to receive the processed output, respectively, so 
\begin{equation}
t_{k}^i= {s_{k}}/{R_{i}},	\: \: k\in\{\text{UL, DL}\}, \: i\in\{\text{HAP, LEO}\}.
\label{eq:t-delay}
\end{equation}

It is important to note that the distance between the swarm of \glspl{uav} and the \gls{leo} satellite is not constant, but varies depending on the elevation angle $\alpha$ as
\begin{equation}
d_{\rm LEO} (\alpha) = \sqrt{R_E^2 \sin{^2\alpha} + h^2 + 2 h R_E} - R_E \sin{\alpha},
\end{equation}
where $R_E$ is the radius of the Earth, and $h$ is the altitude of the satellite orbit~\cite{38811}.
\end{itemize}

With an offloading factor $\eta$, which represents the probability of a frame being offloaded to the edge server (either the HAP or the LEO satellite), the average end-to-end delay is
\begin{equation}
	\Bar{T}=(1-\eta)\Bar{t}_{\rm UAV}+\eta\Bar{t}_i, \: i\in\{\text{HAP, LEO}\}.
	\label{eq:delay}
\end{equation}

\section{Energy Model} 
\label{sec:energy_model}
In this section we present our energy model to characterize the energy consumption of the \glspl{uav}, \gls{hap}, and LEO satellite. Specifically, we distinguish between energy for movement (\cref{sub:e-movement}), data offloading (\cref{sub:e-offloading}), and data processing (\cref{sub:e-processing}), while the overall energy consumption is given in Sec.~\ref{sub:overall_energy_consumption}. Moreover, in~\cref{sub:e-capacity} we describe the energy capacity at the different nodes. 


\subsection{Energy for Movement ($E_{\rm M}$)}
\label{sub:e-movement}
We assume that the LEO satellite and the HAP do not consume energy to move, despite some wind and/or air resistance, i.e., $E^{\rm HAP}_{\rm M}=E^{\rm LEO}_{\rm M}=0$, while instead the UAV incurs a significant energy cost for hovering (i.e., controlling and maintaining the \gls{uav} in the air). The latter is characterized by the model in~\cite{Abeywickrama18}, which is based on empirical data. 
So, the hovering power can be written~as
\begin{equation}
	P^{\rm UAV}_{\rm M} = \sqrt{{\left(m g\right)^3}/({2\pi \xi^2 \psi})},
	\label{eq:power_hovering}
\end{equation}
where $m$ is the mass of the \gls{uav}, $g$ is the gravitational acceleration, $\xi$ is the radius of a propeller, and $\psi$ is the air density.
Finally, the energy consumption for hovering becomes 
\begin{equation}
	E^{\rm UAV}_{\rm M} = P^{\rm UAV}_{\rm M}t_f,
	\label{eq:hovering}
\end{equation}
where $t_f$ represents the total flying time.


\subsection{Energy for Data Offloading ($E_{\rm DO}$)}
\label{sub:e-offloading}
Assuming transmissions in the \gls{mmwave} spectrum, the energy consumption for data offloading to/from the UAV is modeled based on the analysis in~\cite{Pizzo17Optimal}.
The energy consumption depends on the type of antenna that is used to transmit or receive data. We distinguish between \gls{upa} and \gls{car}~antennas. 
\begin{itemize}
	\item \emph{UPA antennas.}
Assuming analog beamforming with $N$ antenna elements, the power consumption to transmit data (i.e., video frames from the UAV to the HAP/LEO, or the processed output from the HAP/LEO to the UAV) is
\begin{equation}
	\resizebox{.8\hsize}{!}{$P^{\rm UPA}_{\rm TX} = \frac{P_{t}}{\zeta} + \Big(N (P_{\rm PS}+P_{\rm HPA}) + P_{\rm DAC} +P_{\rm UC}+P_{\rm C}\Big)$}
	\label{eq10}
\end{equation}
where $P_{t}$ is the transmit power, $\zeta\leq 1$ is the efficiency of the power amplifier, and the second term accounts for the power consumption of the circuitry. Specifically,  $P_{\rm PS}$, $P_{\rm HPA}$, $P_{\rm DAC}$, $P_{\rm UC}$, and $P_{\rm C}$ are the power consumptions of phase shifters, high-power amplifiers, the digital-to-analog converter, the up-conversion stage, and the combiner. In ~\cref{eq10} we neglect the power to generate the precoder, which accounts for no more than $5\%$ of the total power consumption~\cite{Pizzo17Optimal}.

The power consumption to receive data~is
\begin{equation}
	P^{\rm UPA}_{\rm RX} = N (P_{\rm PS}+P_{\rm LNA}) + P_{\rm ADC} + P_{\rm DC}+P_{\rm C},
\end{equation}
where $P_{\rm LNA}$, $P_{\rm ADC}$, and $P_{\rm DC}$ are the power consumptions of low-noise amplifiers, the analog-to-digital converter, and the down-conversion stage.

\item \emph{\gls{car} antennas.}
The power consumption to transmit and receive data is given~by
\begin{subequations}
\begin{alignat}{2}
	&P^{\rm CAR}_{\rm TX} = \frac{P_{t}}{\zeta} + \Big(P_{\rm HPA} + P_{\rm DAC} +P_{\rm UC}+P_{\rm C}\Big);\\
    &P^{\rm CAR}_{\rm RX} = P_{\rm LNA} + P_{\rm ADC} + P_{\rm DC} + P_{\rm C}.
\end{alignat}
\end{subequations}
\end{itemize}

In uplink, the UAV acts as a transmitter and the HAP/LEO as a receiver, while in downlink it is the contrary. Therefore, the total energy consumption for data offloading is
\begin{subequations}
\begin{alignat}{2}
\label{eq:e_do}
	&E_{\text{DO}}^{\text{UAV}} &&= t^{i}_{\rm UL} P^z_{\rm TX}+t^{i}_{\rm DL} P^z_{\rm RX},\\
	&E_{\text{DO}}^{i} &&= t^{i}_{\rm DL} P^z_{\rm TX}+t^{i}_{\rm UL} P^z_{\rm RX} , \: i\in\{\text{HAP, LEO}\},
\end{alignat}
\end{subequations}
where $z\in\{\text{UPA, CAR}\},$ and $t^{i}_{\rm UL}$ and $t^{i}_{\rm DL}$ are the transmission delays to send video frames from the UAV to the HAP/LEO, and the processed output (bounding boxes) from the HAP/LEO to the UAV, respectively, as expressed in~\cref{eq:t-delay}.


\begin{table*}[t!] \label{tab:params1}
\footnotesize
\renewcommand{\arraystretch}{1.3}
\makesavenoteenv{tabular}
\makesavenoteenv{table}
\caption{System parameters.}
    \label{tab:params1}
   \centering
\begin{tabular}{|l|l|l|l|ll|}
\hline
Layer    & UAV                     & HAP                     & LEO                     & \multicolumn{2}{c|}{Other Parameters}                       \\ \hline
GPU Energy efficiency ($\nu$) [GFLOP/J]& [30,90]                      & 200                      & 200                      & \multicolumn{1}{l|}{Computational load ($C_{l}$) [GFLOP]} & \multicolumn{1}{l|}{90}  \\ \hline
Battery capacity ($E_b$) [Wh]     & 130                     & 8000                      & 6000                      & \multicolumn{1}{l|}{Number of UAVs ($n$)}       & \multicolumn{1}{l|}{[5,50]} \\ \hline
Solar panel area ($S$) [m$^2$]    & \multicolumn{1}{l|}{N/A} & \multicolumn{1}{l|}{100} & \multicolumn{1}{l|}{30} & \multicolumn{1}{l|}{Frame rate ($r$) [fps]}    & [1,19]                       \\ \hline
UPA antenna elements ($N$) & \multicolumn{1}{l|}{[4,128]}  & \multicolumn{1}{l|}{64}  & \multicolumn{1}{l|}{N/A$^\dagger$}  & \multicolumn{1}{l|}{Transmitted power ($P_t$) [dBm]}    & 30                       \\ \hline
Effective isotropic radiated power (EIRP) [dB] & \multicolumn{1}{l|}{$10\log_{10}(N)-6$}  & \multicolumn{1}{l|}{12}  & \multicolumn{1}{l|}{32.5}  & \multicolumn{1}{l|}{Hovering power ($P^{\rm UAV}_M$) [W]}    &   $\sim212$                     \\ \hline
Antenna-gain-to-noise-temperature ($G/T$) [dB] & \multicolumn{1}{l|}{$10\log_{10}(N)-31$}  & \multicolumn{1}{l|}{$-13$}  & \multicolumn{1}{l|}{13}  & \multicolumn{1}{l|}{Radius of the Earth ($R_E$) [km]}    & 6371 \\ \hline
Altitude ($h$) [km] & \multicolumn{1}{l|}{0.1}  & \multicolumn{1}{l|}{20}  & \multicolumn{1}{l|}{600}  & \multicolumn{1}{l|}{Boltzmann constant ($K_b$) [J/K]}    &$1.38\cdot10^{-23}$\\ \hline
Computational capacity ($C$) [GFLOP/s] & \multicolumn{1}{l|}{1000}  & \multicolumn{1}{l|}{20000}  & \multicolumn{1}{l|}{20000}  & \multicolumn{1}{l|}{UL payload ($s_{\rm UL}$) [Mb]}    & 3 \\ \hline
Photovoltaic efficiency ($\chi$) & N/A & 0.15  & 0.15 & \multicolumn{1}{l|}{DL payload ($s_{\rm DL}$) [Mb]}    & 0.1 \\ \hline
 Extra-terrestrial solar irradiance density ($I$) [W/m$^2$] & N/A & 600  & 600 &  \multicolumn{1}{l|}{Bandwidth ($B$) [MHz]}  & $B_T/n=400/n$  \\ \hline
 \multicolumn{6}{l}{$^\dagger$According to~\cite{38811}, LEO satellites use a CAR antenna for communication.\vspace{-0.5cm}}  \\ 
\end{tabular}
\end{table*}

\subsection{Energy for Data Processing ($E_{\rm P}$)}
\label{sub:e-processing}
The energy consumption to process (i.e., perform object detection on) a video frame can be calculated as
\begin{equation}
E_{\rm P}^{i} = {C_{l}}/{\nu_{i}}, \: i\in\{\rm UAV, HAP, LEO\},
\end{equation}
where $C_{l}$ is the constant computational load required to process each video frame in GFLOP, and $\nu_{i}$ is the energy efficiency of the \gls{gpu} installed at node $i$, expressed in GFLOP/J.
Given energy and space constraints, only low-cost \glspl{gpu} can be installed onboard the \glspl{uav}, while the \gls{hap} and the \gls{leo} satellite can be equipped with more powerful and advanced, so more efficient, \glspl{gpu}, i.e., $\nu_{\rm LEO}>\nu_{\rm HAP}>\nu_{\text{UAV}}$.
This means that the same processing task will consume more energy if it is performed onboard the UAV than at an edge server.

\subsection{Overall Energy Consumption} 
\label{sub:overall_energy_consumption}
Given an offloading factor $\eta$, the overall average energy consumption of a \gls{uav} is due to the energy for movement/hovering ($E^{\rm UAV}_{\rm M}$), the energy for on board processing ($E_P^{\rm UAV}$, with probability $1-\eta$), and the energy to transmit data frames to the edge server and receive the processed output ($E_{\text{DO}}^{\text{UAV}}$, with probability $\eta$). 
The overall average energy consumption of a HAP/LEO serving a swarm of $n$ UAVs, with probability $\eta$, is due to the energy for data processing ($E_{\rm P}^{i}$) and the energy to receive data frames and transmit the processed output ($E_{\text{DO}}^i$), with $i\in\{\text{HAP, LEO}\}$. 
Therefore: 
\begin{subequations}
	\begin{alignat}{2}
			&E_{\rm UAV} (\eta) = \left[ (1-\eta)E_P^{\rm UAV}+\eta E_{\text{DO}}^{\text{UAV}} \right] rt_f+E^{\rm UAV}_{\rm M};\label{eq:e_uav}\\
			&E_{i} (\eta) = \eta r n \left( E_{\rm P}^{i} + E_{\text{DO}}^i \right) t_f, \: i\in\{\text{HAP, LEO}\}.\label{eq:e_edge}
\end{alignat}
\end{subequations}

\subsection{Energy Capacity ($E_C$)}
\label{sub:e-capacity}
Both the \glspl{uav} and the HAP/LEO edge server have energy constraints, so it is important to model the actual energy capacity, i.e., the energy available at the nodes, in order to understand the impact of the offloading and processing operations on the energy consumption.
For the \gls{uav}, the energy capacity is equal to the energy stored in the batteries, while the \gls{hap} and the \gls{leo} satellite can additionally harvest energy from solar panels, i.e.,
\begin{subequations}
	\begin{alignat}{2}
& E_{C}^{\rm UAV} = E_{b}^{\rm UAV};\\
& E_{C}^{i} = E_{b}^{i} + E_{h}^{i}, \: i\in\{\text{HAP, LEO}\},\label{eq:capacity_edge}
\end{alignat}
\end{subequations}
where $E_{b}$ is the capacity of the batteries, and $E_{ h}$ is the energy harvested from the solar panels. 
Specifically, $E_{ h}$ is a function of the efficiency of the photovoltaic system ($\chi$), the total extra-terrestrial solar irradiance density on the surface of node ($I$), the total area of the solar panels installed on node ($S$)~\cite{solarpanelmodel}, and the flying time ($t_f$), and takes into account the fact that all the energy that could be harvested when the battery is full cannot be stored and is therefore lost.



\begin{figure*}[t!]
    \subfloat[][Stability for $\eta=0$.]
	{
	    \label{fig:local_processing_stability}
\begin{tikzpicture}

\definecolor{darkslategray38}{RGB}{38,38,38}
\definecolor{darkslategray66}{RGB}{66,66,66}
\definecolor{lightgray204}{RGB}{204,204,204}
\definecolor{color3}{RGB}{204,76,2}
\definecolor{color1}{RGB}{254,153,41}
\definecolor{color2}{RGB}{194,230,153}

\pgfplotsset{compat=1.11,
	/pgfplots/ybar legend/.style={
		/pgfplots/legend image code/.code={%
			\draw[##1,/tikz/.cd,yshift=-0.25em]
			(0cm,0cm) rectangle (20pt,0.6em);},
	},
}

\begin{axis}[
width = \textwidth/3.3,
height = 4.5cm,
axis line style={darkslategray38},
legend cell align={left},
legend columns=3,
legend style={
  fill opacity=0.8,
  draw opacity=1,
  text opacity=1,
  at={(0.5,1.1)},
  anchor=north,
  draw=lightgray204,
  /tikz/every even column/.append style={column sep=0.5em}
},
tick align=outside,
unbounded coords=jump,
x grid style={lightgray204},
xlabel=\textcolor{darkslategray38}{Frame rate ($r$) [fps]},
xmajorticks=true,
xmin=-0.5, xmax=5.5,
xtick style={color=darkslategray38},
xtick={0,1,2,3,4,5},
xticklabels={5,10,15,20,25,30},
y grid style={lightgray204},
ylabel=\textcolor{darkslategray38}{Load factor (\(\displaystyle \rho\))},
ymajorticks=true,
ymajorgrids=true,
ymin=0, ymax=3.5,
ytick style={color=darkslategray38}
]
\draw[draw=black,fill=color1,line width=0.08pt] (axis cs:-0.4,0) rectangle (axis cs:0,0.225);

\addplot[ybar legend,ybar,draw=black,fill=color2,line width=0.08pt]
  table[row sep=crcr]{%
	0	0\\
};
\addlegendentry{$\eta=0$}

\draw[draw=black,fill=color1,line width=0.08pt] (axis cs:0.6,0) rectangle (axis cs:1,0.45);
\draw[draw=black,fill=color1,line width=0.08pt] (axis cs:1.6,0) rectangle (axis cs:2,0.675);
\draw[draw=black,fill=color1,line width=0.08pt] (axis cs:2.6,0) rectangle (axis cs:3,0.9);
\draw[draw=black,fill=color1,line width=0.08pt] (axis cs:3.6,0) rectangle (axis cs:4,1.125);
\draw[draw=black,fill=color1,line width=0.08pt] (axis cs:4.6,0) rectangle (axis cs:5,1.35);
\draw[draw=black,fill=color2,line width=0.08pt] (axis cs:-2.77555756156289e-17,0) rectangle (axis cs:0.4,0.45);

\addplot[ybar legend,ybar,draw=black,fill=color1,line width=0.08pt]
  table[row sep=crcr]{%
	0	0\\
};
\addlegendentry{$\eta=0.5$}

\draw[draw=black,fill=color2,line width=0.08pt] (axis cs:1,0) rectangle (axis cs:1.4,0.9);
\draw[draw=black,fill=color2,line width=0.08pt] (axis cs:2,0) rectangle (axis cs:2.4,1.35);
\draw[draw=black,fill=color2,line width=0.08pt] (axis cs:3,0) rectangle (axis cs:3.4,1.8);
\draw[draw=black,fill=color2,line width=0.08pt] (axis cs:4,0) rectangle (axis cs:4.4,2.25);
\draw[draw=black,fill=color2,line width=0.08pt] (axis cs:5,0) rectangle (axis cs:5.4,2.7);
\addplot [line width=1pt, darkslategray66, forget plot]
table {%
-0.2 nan
-0.2 nan
};
\addplot [line width=1pt, darkslategray66, forget plot]
table {%
0.8 nan
0.8 nan
};
\addplot [line width=1pt, darkslategray66, forget plot]
table {%
1.8 nan
1.8 nan
};
\addplot [line width=1pt, darkslategray66, forget plot]
table {%
2.8 nan
2.8 nan
};
\addplot [line width=1pt, darkslategray66, forget plot]
table {%
3.8 nan
3.8 nan
};
\addplot [line width=1pt, darkslategray66, forget plot]
table {%
4.8 nan
4.8 nan
};
\addplot [line width=1pt, darkslategray66, forget plot]
table {%
0.2 nan
0.2 nan
};
\addplot [line width=1pt, darkslategray66, forget plot]
table {%
1.2 nan
1.2 nan
};
\addplot [line width=1pt, darkslategray66, forget plot]
table {%
2.2 nan
2.2 nan
};
\addplot [line width=1pt, darkslategray66, forget plot]
table {%
3.2 nan
3.2 nan
};
\addplot [line width=1pt, darkslategray66, forget plot]
table {%
4.2 nan
4.2 nan
};
\addplot [line width=1pt, darkslategray66, forget plot]
table {%
5.2 nan
5.2 nan
};
\addplot [line width=1pt, black, dashed]
table {%
-0.5 1
5.5 1
};
\addlegendentry{Stability}
\end{axis}

\end{tikzpicture}
	}
   \subfloat[][Stability for $\eta=0.5$.]
	{
		\label{fig:50_offloading_stability}
\begin{tikzpicture}

\definecolor{darkslategray38}{RGB}{38,38,38}
\definecolor{lightgray204}{RGB}{204,204,204}

\begin{axis}[
width = \textwidth/4.5,
height = 4.5cm,
axis line style={darkslategray38},
colorbar,
colorbar style={ytick={0,0.2,0.4,0.6,0.8,1},yticklabels={
  \(\displaystyle {0.0}\),
  \(\displaystyle {0.2}\),
  \(\displaystyle {0.4}\),
  \(\displaystyle {0.6}\),
  \(\displaystyle {0.8}\),
  \(\displaystyle {1.0}\)
},ylabel={Load factor ($\rho$)}},
colormap={mymap}{[1pt]
  rgb(0pt)=(1,1,1);
  rgb(1pt)=(0.992156862745098,0.6,0.16078431372549)
},
point meta max=1,
point meta min=0,
tick align=outside,
x grid style={lightgray204},
xlabel=\textcolor{darkslategray38}{Frame rate ($r$) [fps]},
xmajorticks=true,
xmin=0, xmax=4,
xtick style={color=darkslategray38},
xtick={0.5,1.5,2.5,3.5},
xticklabels={5,10,15,20},
y dir=reverse,
y grid style={lightgray204},
ylabel=\textcolor{darkslategray38}{Number of UAVs ($n$)},
ymajorticks=true,
ymin=0, ymax=6,
ytick style={color=darkslategray38},
ytick={0.5,1.5,2.5,3.5,4.5,5.5},
yticklabels={5,10,15,20,25,30}
]
\addplot graphics [includegraphics cmd=\pgfimage,xmin=0, xmax=4, ymin=6, ymax=0] {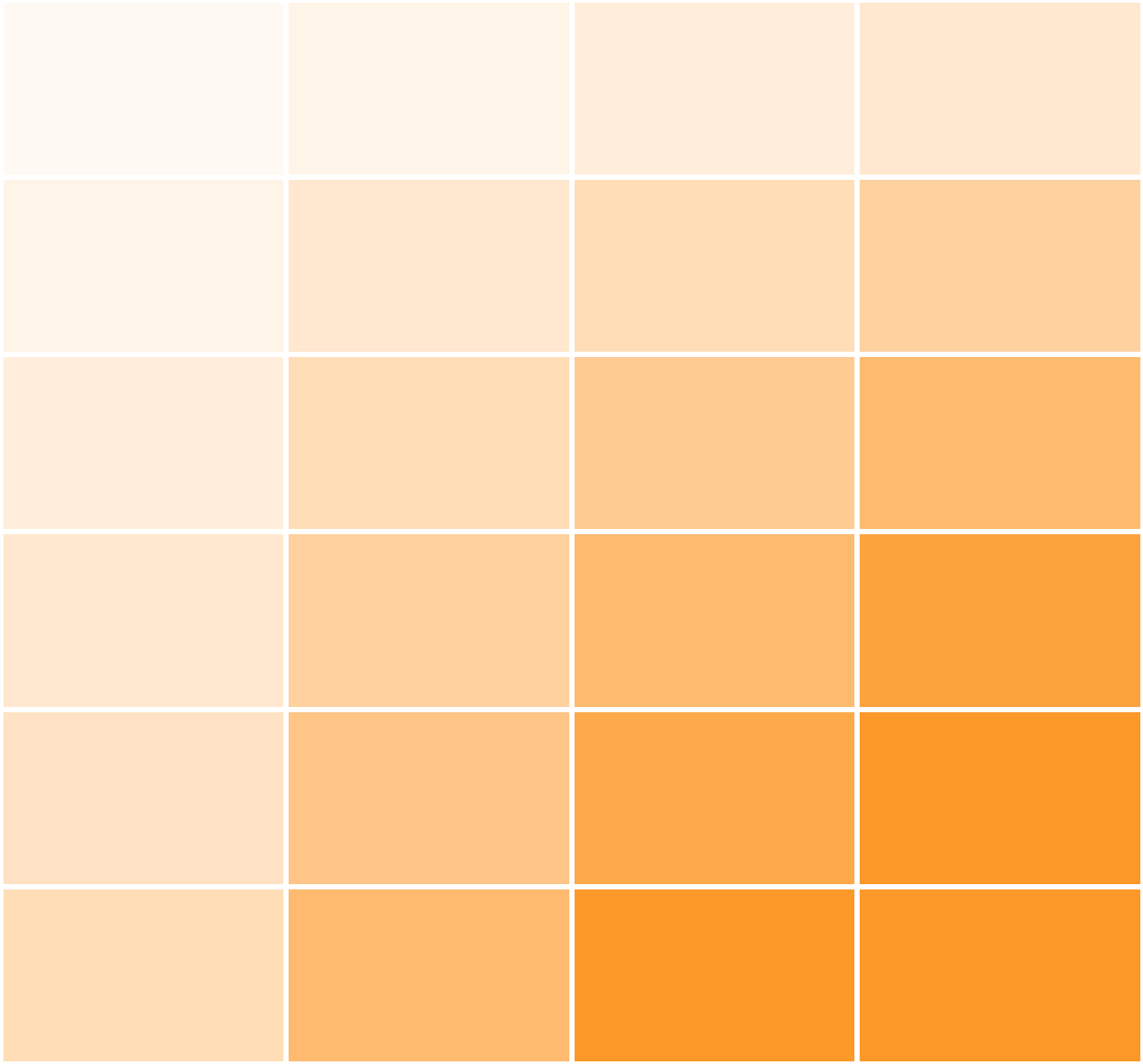};
\draw (axis cs:0.5,0.5) node[
  scale=0.5,
  text=darkslategray38,
  rotate=0.0
]{0.056};
\draw (axis cs:1.5,0.5) node[
  scale=0.5,
  text=darkslategray38,
  rotate=0.0
]{0.11};
\draw (axis cs:2.5,0.5) node[
  scale=0.5,
  text=darkslategray38,
  rotate=0.0
]{0.17};
\draw (axis cs:3.5,0.5) node[
  scale=0.5,
  text=darkslategray38,
  rotate=0.0
]{0.23};
\draw (axis cs:0.5,1.5) node[
  scale=0.5,
  text=darkslategray38,
  rotate=0.0
]{0.11};
\draw (axis cs:1.5,1.5) node[
  scale=0.5,
  text=darkslategray38,
  rotate=0.0
]{0.23};
\draw (axis cs:2.5,1.5) node[
  scale=0.5,
  text=darkslategray38,
  rotate=0.0
]{0.34};
\draw (axis cs:3.5,1.5) node[
  scale=0.5,
  text=darkslategray38,
  rotate=0.0
]{0.45};
\draw (axis cs:0.5,2.5) node[
  scale=0.5,
  text=darkslategray38,
  rotate=0.0
]{0.17};
\draw (axis cs:1.5,2.5) node[
  scale=0.5,
  text=darkslategray38,
  rotate=0.0
]{0.34};
\draw (axis cs:2.5,2.5) node[
  scale=0.5,
  text=white,
  rotate=0.0
]{0.51};
\draw (axis cs:3.5,2.5) node[
  scale=0.5,
  text=white,
  rotate=0.0
]{0.67};
\draw (axis cs:0.5,3.5) node[
  scale=0.5,
  text=darkslategray38,
  rotate=0.0
]{0.23};
\draw (axis cs:1.5,3.5) node[
  scale=0.5,
  text=darkslategray38,
  rotate=0.0
]{0.45};
\draw (axis cs:2.5,3.5) node[
  scale=0.5,
  text=white,
  rotate=0.0
]{0.67};
\draw (axis cs:3.5,3.5) node[
  scale=0.5,
  text=white,
  rotate=0.0
]{0.9};
\draw (axis cs:0.5,4.5) node[
  scale=0.5,
  text=darkslategray38,
  rotate=0.0
]{0.28};
\draw (axis cs:1.5,4.5) node[
  scale=0.5,
  text=white,
  rotate=0.0
]{0.56};
\draw (axis cs:2.5,4.5) node[
  scale=0.5,
  text=white,
  rotate=0.0
]{0.84};
\draw (axis cs:3.5,4.5) node[
  scale=0.5,
  text=white,
  rotate=0.0
]{1.1};
\draw (axis cs:0.5,5.5) node[
  scale=0.5,
  text=darkslategray38,
  rotate=0.0
]{0.34};
\draw (axis cs:1.5,5.5) node[
  scale=0.5,
  text=white,
  rotate=0.0
]{0.67};
\draw (axis cs:2.5,5.5) node[
  scale=0.5,
  text=white,
  rotate=0.0
]{1};
\draw (axis cs:3.5,5.5) node[
  scale=0.5,
  text=white,
  rotate=0.0
]{1.3};
\end{axis}

\end{tikzpicture}
	}
	\subfloat[][Stability for $\eta=1$.]
	{
            \label{fig:100_offloading_stability}
\begin{tikzpicture}

\definecolor{darkslategray38}{RGB}{38,38,38}
\definecolor{lightgray204}{RGB}{204,204,204}

\begin{axis}[
width = \textwidth/4.5,
height = 4.5cm,
axis line style={darkslategray38},
colorbar,
colorbar style={ytick={0,0.2,0.4,0.6,0.8,1},yticklabels={
  \(\displaystyle {0.0}\),
  \(\displaystyle {0.2}\),
  \(\displaystyle {0.4}\),
  \(\displaystyle {0.6}\),
  \(\displaystyle {0.8}\),
  \(\displaystyle {1.0}\)
},ylabel={Load factor ($\rho$)}},
colormap={mymap}{[1pt]
  rgb(0pt)=(1,1,1);
  rgb(1pt)=(0.8,0.298039215686275,0.00784313725490196)
},
point meta max=1,
point meta min=0,
tick align=outside,
x grid style={lightgray204},
xlabel=\textcolor{darkslategray38}{Frame rate ($r$) [fps]},
xmajorticks=true,
xmin=0, xmax=4,
xtick style={color=darkslategray38},
xtick={0.5,1.5,2.5,3.5},
xticklabels={5,10,15,20},
y dir=reverse,
y grid style={lightgray204},
ylabel=\textcolor{darkslategray38}{Number of UAVs ($n$)},
ymajorticks=true,
ymin=0, ymax=6,
ytick style={color=darkslategray38},
ytick={0.5,1.5,2.5,3.5,4.5,5.5},
yticklabels={5,10,15,20,25,30}
]
\addplot graphics [includegraphics cmd=\pgfimage,xmin=0, xmax=4, ymin=6, ymax=0] {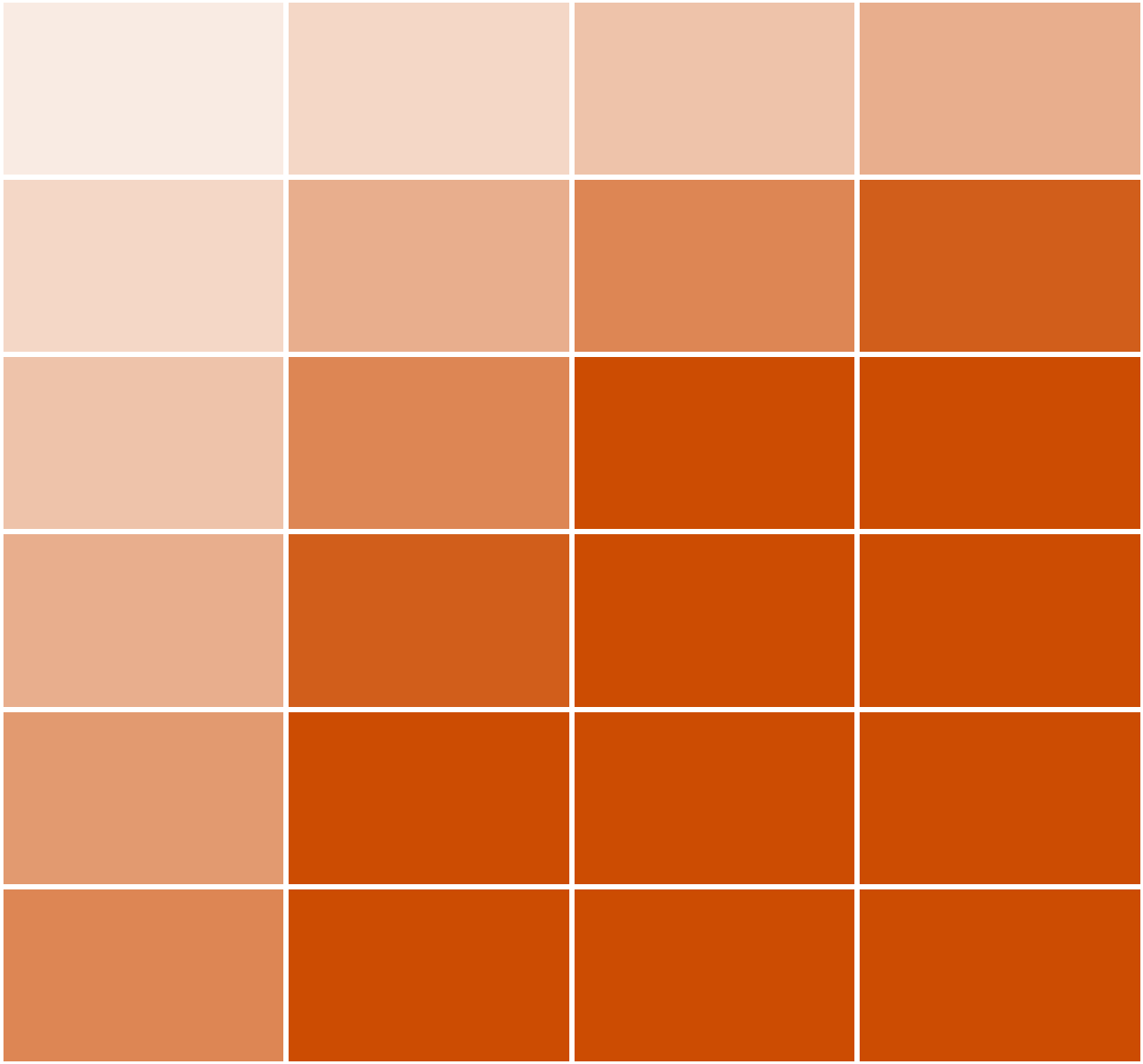};
\draw (axis cs:0.5,0.5) node[
  scale=0.5,
  text=darkslategray38,
  rotate=0.0
]{0.11};
\draw (axis cs:1.5,0.5) node[
  scale=0.5,
  text=darkslategray38,
  rotate=0.0
]{0.23};
\draw (axis cs:2.5,0.5) node[
  scale=0.5,
  text=darkslategray38,
  rotate=0.0
]{0.34};
\draw (axis cs:3.5,0.5) node[
  scale=0.5,
  text=white,
  rotate=0.0
]{0.45};
\draw (axis cs:0.5,1.5) node[
  scale=0.5,
  text=darkslategray38,
  rotate=0.0
]{0.23};
\draw (axis cs:1.5,1.5) node[
  scale=0.5,
  text=white,
  rotate=0.0
]{0.45};
\draw (axis cs:2.5,1.5) node[
  scale=0.5,
  text=white,
  rotate=0.0
]{0.67};
\draw (axis cs:3.5,1.5) node[
  scale=0.5,
  text=white,
  rotate=0.0
]{0.9};
\draw (axis cs:0.5,2.5) node[
  scale=0.5,
  text=darkslategray38,
  rotate=0.0
]{0.34};
\draw (axis cs:1.5,2.5) node[
  scale=0.5,
  text=white,
  rotate=0.0
]{0.67};
\draw (axis cs:2.5,2.5) node[
  scale=0.5,
  text=white,
  rotate=0.0
]{1};
\draw (axis cs:3.5,2.5) node[
  scale=0.5,
  text=white,
  rotate=0.0
]{1.3};
\draw (axis cs:0.5,3.5) node[
  scale=0.5,
  text=white,
  rotate=0.0
]{0.45};
\draw (axis cs:1.5,3.5) node[
  scale=0.5,
  text=white,
  rotate=0.0
]{0.9};
\draw (axis cs:2.5,3.5) node[
  scale=0.5,
  text=white,
  rotate=0.0
]{1.3};
\draw (axis cs:3.5,3.5) node[
  scale=0.5,
  text=white,
  rotate=0.0
]{1.8};
\draw (axis cs:0.5,4.5) node[
  scale=0.5,
  text=white,
  rotate=0.0
]{0.56};
\draw (axis cs:1.5,4.5) node[
  scale=0.5,
  text=white,
  rotate=0.0
]{1.1};
\draw (axis cs:2.5,4.5) node[
  scale=0.5,
  text=white,
  rotate=0.0
]{1.7};
\draw (axis cs:3.5,4.5) node[
  scale=0.5,
  text=white,
  rotate=0.0
]{2.2};
\draw (axis cs:0.5,5.5) node[
  scale=0.5,
  text=white,
  rotate=0.0
]{0.67};
\draw (axis cs:1.5,5.5) node[
  scale=0.5,
  text=white,
  rotate=0.0
]{1.3};
\draw (axis cs:2.5,5.5) node[
  scale=0.5,
  text=white,
  rotate=0.0
]{2};
\draw (axis cs:3.5,5.5) node[
  scale=0.5,
  text=white,
  rotate=0.0
]{2.7};
\end{axis}

\end{tikzpicture}
	}
    \caption{Stability of the D/M/1 queues as a function of the frame rate $r$, the number of UAVs $n$, and the offloading factor $\eta$.\vspace{-0.5cm}}
    \label{fig:stability}
\end{figure*}
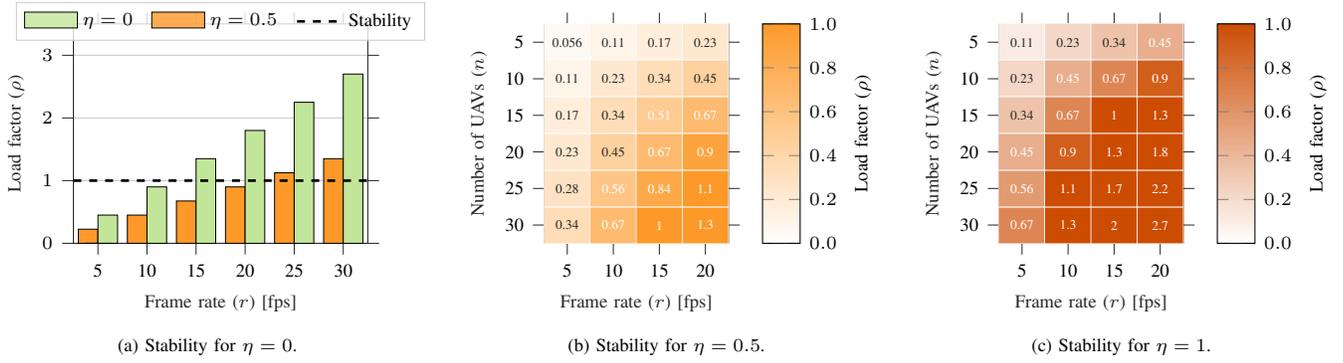

\begin{figure*}[t!]
    \subfloat[][$\nu_{\rm UAV}=50$~GFLOP/J, $n=20$, and $r=10$ fps.\vspace{-0.5cm}]
	{
		\label{fig:HAP_N}
        \vspace{-0.33cm}
\begin{tikzpicture}
\pgfplotsset{every tick label/.append style={font=\scriptsize}}

\pgfplotsset{compat=1.11,
	/pgfplots/ybar legend/.style={
		/pgfplots/legend image code/.code={%
			\draw[##1,/tikz/.cd,yshift=-0.25em]
			(0cm,0cm) rectangle (20pt,0.6em);},
	},
}

\definecolor{darkgray176}{RGB}{176,176,176}
\definecolor{darkslategray66}{RGB}{66,66,66}
\definecolor{lightgray204}{RGB}{204,204,204}
\definecolor{color1}{RGB}{204,76,2}
\definecolor{color2}{RGB}{254,153,41}
\definecolor{color3}{RGB}{194,230,153}

\begin{axis}[
width = \textwidth/2,
height = 4cm,
legend cell align={left},
legend style={legend cell align=left,
              align=center,
              draw=white!15!black,
              at={(0.5, 1)},
              anchor=center,
              /tikz/every even column/.append style={column sep=1em}},
legend columns=3,
tick pos=both,
unbounded coords=jump,
x grid style={darkgray176},
xlabel={Number of antenna elements (\(\displaystyle N\))},
xmin=-0.5, xmax=5.5,
xtick style={color=black},
xtick={0,1,2,3,4,5},
xticklabels={4,8,16,32,64,128},
y grid style={darkgray176},
ylabel={Average UAV autonomy ($\Bar{\kappa}$) [\%]},
ymajorgrids,
ymin=80, ymax=100,
ytick style={color=black},
]
\draw[draw=black,fill=color1,line width=0.08pt] (axis cs:-0.4,0) rectangle (axis cs:-0.133333333333333,95.487473380216);
\addlegendimage{ybar,ybar legend,draw=black,fill=color1,line width=0.08pt}
\addlegendentry{$\eta=1$}

\draw[draw=black,fill=color1,line width=0.08pt] (axis cs:0.6,0) rectangle (axis cs:0.866666666666667,96.6834976041784);
\draw[draw=black,fill=color1,line width=0.08pt] (axis cs:1.6,0) rectangle (axis cs:1.86666666666667,97.1100412485383);
\draw[draw=black,fill=color1,line width=0.08pt] (axis cs:2.6,0) rectangle (axis cs:2.86666666666667,96.9937637378682);
\draw[draw=black,fill=color1,line width=0.08pt] (axis cs:3.6,0) rectangle (axis cs:3.86666666666667,96.3559027380719);
\draw[draw=black,fill=color1,line width=0.08pt] (axis cs:4.6,0) rectangle (axis cs:4.86666666666667,95.0487752787233);
\draw[draw=black,fill=color2,line width=0.08pt] (axis cs:-0.133333333333333,0) rectangle (axis cs:0.133333333333333,93.8069258219684);
\addlegendimage{ybar,ybar legend,draw=black,fill=color2,line width=0.08pt}
\addlegendentry{$\eta=0.5$}

\draw[draw=black,fill=color2,line width=0.08pt] (axis cs:0.866666666666667,0) rectangle (axis cs:1.13333333333333,94.3804186996472);
\draw[draw=black,fill=color2,line width=0.08pt] (axis cs:1.86666666666667,0) rectangle (axis cs:2.13333333333333,94.5831929752798);
\draw[draw=black,fill=color2,line width=0.08pt] (axis cs:2.86666666666667,0) rectangle (axis cs:3.13333333333333,94.5280065517338);
\draw[draw=black,fill=color2,line width=0.08pt] (axis cs:3.86666666666667,0) rectangle (axis cs:4.13333333333333,94.2240606911474);
\draw[draw=black,fill=color2,line width=0.08pt] (axis cs:4.86666666666667,0) rectangle (axis cs:5.13333333333333,93.5947337521928);
\draw[draw=black,fill=color3,line width=0.08pt] (axis cs:0.133333333333333,0) rectangle (axis cs:0.4,92.1845093251185);
\addlegendimage{ybar,ybar legend,draw=black,fill=color3,line width=0.08pt}
\addlegendentry{$\eta=0$}

\draw[draw=black,fill=color3,line width=0.08pt] (axis cs:1.13333333333333,0) rectangle (axis cs:1.4,92.1845093251185);
\draw[draw=black,fill=color3,line width=0.08pt] (axis cs:2.13333333333333,0) rectangle (axis cs:2.4,92.1845093251185);
\draw[draw=black,fill=color3,line width=0.08pt] (axis cs:3.13333333333333,0) rectangle (axis cs:3.4,92.1845093251185);
\draw[draw=black,fill=color3,line width=0.08pt] (axis cs:4.13333333333333,0) rectangle (axis cs:4.4,92.1845093251185);
\draw[draw=black,fill=color3,line width=0.08pt] (axis cs:5.13333333333333,0) rectangle (axis cs:5.4,92.1845093251185);
\addplot [line width=0.864pt, darkslategray66, forget plot]
table {%
-0.266666666666667 nan
-0.266666666666667 nan
};
\addplot [line width=0.864pt, darkslategray66, forget plot]
table {%
0.733333333333333 nan
0.733333333333333 nan
};
\addplot [line width=0.864pt, darkslategray66, forget plot]
table {%
1.73333333333333 nan
1.73333333333333 nan
};
\addplot [line width=0.864pt, darkslategray66, forget plot]
table {%
2.73333333333333 nan
2.73333333333333 nan
};
\addplot [line width=0.864pt, darkslategray66, forget plot]
table {%
3.73333333333333 nan
3.73333333333333 nan
};
\addplot [line width=0.864pt, darkslategray66, forget plot]
table {%
4.73333333333333 nan
4.73333333333333 nan
};
\addplot [line width=0.864pt, darkslategray66, forget plot]
table {%
0 nan
0 nan
};
\addplot [line width=0.864pt, darkslategray66, forget plot]
table {%
1 nan
1 nan
};
\addplot [line width=0.864pt, darkslategray66, forget plot]
table {%
2 nan
2 nan
};
\addplot [line width=0.864pt, darkslategray66, forget plot]
table {%
3 nan
3 nan
};
\addplot [line width=0.864pt, darkslategray66, forget plot]
table {%
4 nan
4 nan
};
\addplot [line width=0.864pt, darkslategray66, forget plot]
table {%
5 nan
5 nan
};
\addplot [line width=0.864pt, darkslategray66, forget plot]
table {%
0.266666666666667 nan
0.266666666666667 nan
};
\addplot [line width=0.864pt, darkslategray66, forget plot]
table {%
1.26666666666667 nan
1.26666666666667 nan
};
\addplot [line width=0.864pt, darkslategray66, forget plot]
table {%
2.26666666666667 nan
2.26666666666667 nan
};
\addplot [line width=0.864pt, darkslategray66, forget plot]
table {%
3.26666666666667 nan
3.26666666666667 nan
};
\addplot [line width=0.864pt, darkslategray66, forget plot]
table {%
4.26666666666667 nan
4.26666666666667 nan
};
\addplot [line width=0.864pt, darkslategray66, forget plot]
table {%
5.26666666666667 nan
5.26666666666667 nan
};
\end{axis}

\end{tikzpicture}
	}
 \hfill
    \subfloat[][$N=8$, $\nu_{\rm UAV}=50$~GFLOP/J, and $n=15$.\vspace{-0.5cm}]
	{
		\label{fig:HAP_r}
        \vspace{-0.33cm}
\begin{tikzpicture}
\pgfplotsset{compat=1.11,
	/pgfplots/ybar legend/.style={
		/pgfplots/legend image code/.code={%
			\draw[##1,/tikz/.cd,yshift=-0.25em]
			(0cm,0cm) rectangle (20pt,0.6em);},
	},
}

\definecolor{darkgray176}{RGB}{176,176,176}
\definecolor{lightgray204}{RGB}{204,204,204}
\definecolor{color1}{RGB}{204,76,2}
\definecolor{color2}{RGB}{254,153,41}
\definecolor{color3}{RGB}{194,230,153}

\begin{axis}[
width = \textwidth/2,
height = 4cm,
tick pos=both,
legend cell align={left},
legend style={legend cell align=left,
              align=center,
              draw=white!15!black,
              at={(0.5, 1)},
              anchor=center,
              /tikz/every even column/.append style={column sep=1em}},
legend columns=4,
unbounded coords=jump,
x grid style={darkgray176},
xlabel={Frame rate ($r$) [fps]},
xmajorgrids,
xmin=-0.5, xmax=6,
xtick style={color=black},
xtick={0,1,2,3,4,5,6},
xticklabels={1,4,7,10,13,16,19},
y grid style={darkgray176},
ylabel={Average delay ($\Bar{T}$) [s]},
ymajorgrids,
ymin=0, ymax=0.6,
ytick style={color=black}
]

\addplot [line width=1pt, color1, mark=*]
table {%
0 0.262612496577725
0 0.262612496577725
};
\addlegendentry{$\eta=1$}

\addplot [line width=1pt, color2, mark=*]
table {%
0 0.176306247470604
0 0.176306247470604
};
\addlegendentry{$\eta=0.5$}

\addplot [line width=1pt, color3, mark=*]
table {%
0 0.0900013453239957
0 0.0900013453239957
};
\addlegendentry{$\eta=0$}


\addplot [line width=1pt, color1, forget plot]
table {%
0 0.262612496577725
1 0.262738544103007
2 0.263561440145006
3 0.26599946960881
4 0.277297157441452
4 0.8
};
\addplot [line width=1pt, color1, forget plot]
table {%
0 nan
0 nan
};
\addplot [line width=1pt, color1, forget plot]
table {%
1 nan
1 nan
};
\addplot [line width=1pt, color1, forget plot]
table {%
2 nan
2 nan
};
\addplot [line width=1pt, color1, forget plot]
table {%
3 nan
3 nan
};
\addplot [line width=1pt, color1, forget plot]
table {%
4 nan
4 nan
};
\addplot [line width=1pt, color1, forget plot]
table {%
5 nan
5 nan
};
\addplot [line width=1pt, color1, forget plot]
table {%
6 nan
6 nan
};
\addplot [draw=color1, fill=color1, mark=*, only marks, forget plot]
table{%
x  y
0 0.262612496577725
1 0.262738544103007
2 0.263561440145006
3 0.26599946960881
4 0.277297157441452
};
\addplot [line width=1pt, color2, forget plot]
table {%
0 0.176306247470604
1 0.176486153067508
2 0.178651307195473
3 0.184513094526117
4 0.196389507339441
5 0.221673561335852
6 0.296173264058662
};
\addplot [line width=1pt, color2, forget plot]
table {%
0 nan
0 nan
};
\addplot [line width=1pt, color2, forget plot]
table {%
1 nan
1 nan
};
\addplot [line width=1pt, color2, forget plot]
table {%
2 nan
2 nan
};
\addplot [line width=1pt, color2, forget plot]
table {%
3 nan
3 nan
};
\addplot [line width=1pt, color2, forget plot]
table {%
4 nan
4 nan
};
\addplot [line width=1pt, color2, forget plot]
table {%
5 nan
5 nan
};
\addplot [line width=1pt, color2, forget plot]
table {%
6 nan
6 nan
};
\addplot [draw=color2, fill=color2, mark=*, only marks, forget plot]
table{%
x  y
0 0.176306247470604
1 0.176486153067508
2 0.178651307195473
3 0.184513094526117
4 0.196389507339441
5 0.221673561335852
6 0.296173264058662
};

\addplot [line width=1pt, color3, forget plot]
table {%
0 0.0900013453239957
1 0.0975087565964022
2 0.141717068687348
3 0.466079347993467
3 0.8
};
\addplot [line width=1pt, color3, forget plot]
table {%
0 nan
0 nan
};
\addplot [line width=1pt, color3, forget plot]
table {%
1 nan
1 nan
};
\addplot [line width=1pt, color3, forget plot]
table {%
2 nan
2 nan
};
\addplot [line width=1pt, color3, forget plot]
table {%
3 nan
3 nan
};
\addplot [line width=1pt, color3, forget plot]
table {%
4 nan
4 nan
};
\addplot [line width=1pt, color3, forget plot]
table {%
5 nan
5 nan
};
\addplot [line width=1pt, color3, forget plot]
table {%
6 nan
6 nan
};
\addplot [draw=color3, fill=color3, mark=*, only marks, forget plot]
table{%
x  y
0 0.0900013453239957
1 0.0975087565964022
2 0.141717068687348
3 0.466079347993467
};

\usetikzlibrary {patterns.meta}
\pgfdeclarepattern{
  name=hatch,
  parameters={\hatchsize,\hatchangle,\hatchlinewidth},
  bottom left={\pgfpoint{-.1pt}{-.1pt}},
  top right={\pgfpoint{\hatchsize+.1pt}{\hatchsize+.1pt}},
  tile size={\pgfpoint{\hatchsize}{\hatchsize}},
  tile transformation={\pgftransformrotate{\hatchangle}},
  code={
    \pgfsetlinewidth{\hatchlinewidth}
    \pgfpathmoveto{\pgfpoint{-.1pt}{-.1pt}}
    \pgfpathlineto{\pgfpoint{\hatchsize+.1pt}{\hatchsize+.1pt}}
    \pgfpathmoveto{\pgfpoint{-.1pt}{\hatchsize+.1pt}}
    \pgfusepath{stroke}
  }
}

\tikzset{
  hatch size/.store in=\hatchsize,
  hatch angle/.store in=\hatchangle,
  hatch line width/.store in=\hatchlinewidth,
  hatch size=5pt,
  hatch angle=0pt,
  hatch line width=.5pt,
}


\end{axis}
\end{tikzpicture}
	}
 \vskip 0.1cm
    \subfloat[][$N=8$, and $n=20$, and $r=10$ fps.]
	{
        \label{fig:HAP_eff}
\begin{tikzpicture}

\definecolor{darkgray176}{RGB}{176,176,176}
\definecolor{darkslategray66}{RGB}{66,66,66}
\definecolor{color1}{RGB}{204,76,2}
\definecolor{color2}{RGB}{254,153,41}
\definecolor{color3}{RGB}{194,230,153}

\begin{axis}[
width = \textwidth/2,
height = 4cm,
tick pos=both,
unbounded coords=jump,
x grid style={darkgray176},
xlabel={UAV energy efficiency (\(\displaystyle \nu_{\rm UAV}\)) [GFLOP/J]},
xmin=-0.5, xmax=3.5,
xtick style={color=black},
xtick={0,1,2,3},
xticklabels={30,50,70,90},
y grid style={darkgray176},
ylabel={Average UAV autonomy ($\Bar{\kappa}$) [\%]},
ymajorgrids,
ymin=80, ymax=100,
ytick style={color=black}
]
\draw[draw=black,fill=color1,line width=0.08pt] (axis cs:-0.4,0) rectangle (axis cs:-0.133333333333333,96.6834976041784);

\draw[draw=black,fill=color1,line width=0.08pt] (axis cs:0.6,0) rectangle (axis cs:0.866666666666667,96.6834976041784);
\draw[draw=black,fill=color1,line width=0.08pt] (axis cs:1.6,0) rectangle (axis cs:1.86666666666667,96.6834976041784);
\draw[draw=black,fill=color1,line width=0.08pt] (axis cs:2.6,0) rectangle (axis cs:2.86666666666667,96.6834976041784);
\draw[draw=black,fill=color2,line width=0.08pt] (axis cs:-0.133333333333333,0) rectangle (axis cs:0.133333333333333,91.9284832070247);

\draw[draw=black,fill=color2,line width=0.08pt] (axis cs:0.866666666666667,0) rectangle (axis cs:1.13333333333333,94.3804186996472);
\draw[draw=black,fill=color2,line width=0.08pt] (axis cs:1.86666666666667,0) rectangle (axis cs:2.13333333333333,95.4717510992468);
\draw[draw=black,fill=color2,line width=0.08pt] (axis cs:2.86666666666667,0) rectangle (axis cs:3.13333333333333,96.0890228770419);
\draw[draw=black,fill=color3,line width=0.08pt] (axis cs:0.133333333333333,0) rectangle (axis cs:0.4,87.6192592985337);

\draw[draw=black,fill=color3,line width=0.08pt] (axis cs:1.13333333333333,0) rectangle (axis cs:1.4,92.1845093251185);
\draw[draw=black,fill=color3,line width=0.08pt] (axis cs:2.13333333333333,0) rectangle (axis cs:2.4,94.2900025703269);
\draw[draw=black,fill=color3,line width=0.08pt] (axis cs:3.13333333333333,0) rectangle (axis cs:3.4,95.5018139304873);
\addplot [line width=0.864pt, darkslategray66]
table {%
-0.266666666666667 nan
-0.266666666666667 nan
};
\addplot [line width=0.864pt, darkslategray66]
table {%
0.733333333333333 nan
0.733333333333333 nan
};
\addplot [line width=0.864pt, darkslategray66]
table {%
1.73333333333333 nan
1.73333333333333 nan
};
\addplot [line width=0.864pt, darkslategray66]
table {%
2.73333333333333 nan
2.73333333333333 nan
};
\addplot [line width=0.864pt, darkslategray66]
table {%
0 nan
0 nan
};
\addplot [line width=0.864pt, darkslategray66]
table {%
1 nan
1 nan
};
\addplot [line width=0.864pt, darkslategray66]
table {%
2 nan
2 nan
};
\addplot [line width=0.864pt, darkslategray66]
table {%
3 nan
3 nan
};
\addplot [line width=0.864pt, darkslategray66]
table {%
0.266666666666667 nan
0.266666666666667 nan
};
\addplot [line width=0.864pt, darkslategray66]
table {%
1.26666666666667 nan
1.26666666666667 nan
};
\addplot [line width=0.864pt, darkslategray66]
table {%
2.26666666666667 nan
2.26666666666667 nan
};
\addplot [line width=0.864pt, darkslategray66]
table {%
3.26666666666667 nan
3.26666666666667 nan
};
\end{axis}

\end{tikzpicture}
	}
 \hfill
      \subfloat[][$N=8$, $\nu_{\rm UAV}=50$~GFLOP/J, and $r=10$ fps.]
	{
		\label{fig:HAP_nUAVs}
\begin{tikzpicture}

\definecolor{darkgray176}{RGB}{176,176,176}
\definecolor{lightgray204}{RGB}{204,204,204}
\definecolor{color1}{RGB}{204,76,2}
\definecolor{color2}{RGB}{254,153,41}
\definecolor{color3}{RGB}{194,230,153}

\begin{axis}[
width = \textwidth/2,
height = 4cm,
tick pos=both,
unbounded coords=jump,
x grid style={darkgray176},
xlabel={Number of UAVs (\(\displaystyle n\))},
xmajorgrids,
xmin=-0.5, xmax=9.5,
xtick style={color=black},
xtick={0,1,2,3,4,5,6,7,8,9},
xticklabels={5,10,15,20,25,30,35,40,45,50},
y grid style={darkgray176},
ylabel={Average delay ($\Bar{T}$) [s]},
ymajorgrids,
ymin=0.157171685977446, ymax=0.480789236660897,
ytick style={color=black}
]
\addplot [line width=1pt, color1, forget plot]
table {%
0 0.237703349163811
1 0.251137659771605
2 0.26599946960881
3 0.293347415283743
3 0.8
};
\addplot [line width=1pt, color1, forget plot]
table {%
0 nan
0 nan
};
\addplot [line width=1pt, color1, forget plot]
table {%
1 nan
1 nan
};
\addplot [line width=1pt, color1, forget plot]
table {%
2 nan
2 nan
};
\addplot [line width=1pt, color1, forget plot]
table {%
3 nan
3 nan
};
\addplot [line width=1pt, color1, forget plot]
table {%
4 nan
4 nan
};
\addplot [line width=1pt, color1, forget plot]
table {%
5 nan
5 nan
};
\addplot [line width=1pt, color1, forget plot]
table {%
6 nan
6 nan
};
\addplot [line width=1pt, color1, forget plot]
table {%
7 nan
7 nan
};
\addplot [line width=1pt, color1, forget plot]
table {%
8 nan
8 nan
};
\addplot [line width=1pt, color1, forget plot]
table {%
9 nan
9 nan
};
\addplot [draw=color1, fill=color1, mark=*, only marks]
table{%
x  y
0 0.237703349163811
1 0.251137659771605
2 0.26599946960881
3 0.293347415283743
};
\addplot [line width=1pt, color2, forget plot]
table {%
0 0.171881574644876
1 0.178252070691306
2 0.184513094526117
3 0.190732478633152
4 0.197000048563659
5 0.203511398361525
6 0.210854388789648
7 0.222218658932818
7 0.480789236660897
};
\addplot [line width=1pt, color2, forget plot]
table {%
0 nan
0 nan
};
\addplot [line width=1pt, color2, forget plot]
table {%
1 nan
1 nan
};
\addplot [line width=1pt, color2, forget plot]
table {%
2 nan
2 nan
};
\addplot [line width=1pt, color2, forget plot]
table {%
3 nan
3 nan
};
\addplot [line width=1pt, color2, forget plot]
table {%
4 nan
4 nan
};
\addplot [line width=1pt, color2, forget plot]
table {%
5 nan
5 nan
};
\addplot [line width=1pt, color2, forget plot]
table {%
6 nan
6 nan
};
\addplot [line width=1pt, color2, forget plot]
table {%
7 nan
7 nan
};
\addplot [line width=1pt, color2, forget plot]
table {%
8 nan
8 nan
};
\addplot [line width=1pt, color2, forget plot]
table {%
9 nan
9 nan
};
\addplot [draw=color2, fill=color2, mark=*, only marks]
table{%
x  y
0 0.171881574644876
1 0.178252070691306
2 0.184513094526117
3 0.190732478633152
4 0.197000048563659
5 0.203511398361525
6 0.210854388789648
7 0.222218658932818
};
\addplot [line width=1pt, color3, forget plot]
table {%
0 0.466079347993467
1 0.466079347993467
2 0.466079347993467
3 0.466079347993467
4 0.466079347993467
5 0.466079347993467
6 0.466079347993467
7 0.466079347993467
8 0.466079347993467
9 0.466079347993467
};
\addplot [line width=1pt, color3, forget plot]
table {%
0 nan
0 nan
};
\addplot [line width=1pt, color3, forget plot]
table {%
1 nan
1 nan
};
\addplot [line width=1pt, color3, forget plot]
table {%
2 nan
2 nan
};
\addplot [line width=1pt, color3, forget plot]
table {%
3 nan
3 nan
};
\addplot [line width=1pt, color3, forget plot]
table {%
4 nan
4 nan
};
\addplot [line width=1pt, color3, forget plot]
table {%
5 nan
5 nan
};
\addplot [line width=1pt, color3, forget plot]
table {%
6 nan
6 nan
};
\addplot [line width=1pt, color3, forget plot]
table {%
7 nan
7 nan
};
\addplot [line width=1pt, color3, forget plot]
table {%
8 nan
8 nan
};
\addplot [line width=1pt, color3, forget plot]
table {%
9 nan
9 nan
};
\addplot [draw=color3, fill=color3, mark=*, only marks]
table{%
x  y
0 0.466079347993467
1 0.466079347993467
2 0.466079347993467
3 0.466079347993467
4 0.466079347993467
5 0.466079347993467
6 0.466079347993467
7 0.466079347993467
8 0.466079347993467
9 0.466079347993467
};

\tikzset{
  hatch size/.store in=\hatchsize,
  hatch angle/.store in=\hatchangle,
  hatch line width/.store in=\hatchlinewidth,
  hatch size=5pt,
  hatch angle=0pt,
  hatch line width=.5pt,
}


\end{axis}

\end{tikzpicture}
	}
    \caption{HAP-assisted edge computing. Average UAV autonomy $\bar{\kappa}$ (left) and average delay $\bar{T}$ (right) vs. $\eta$, as a function of the number of antenna elements $N$ at the UAV, the energy efficiency $\nu_{\rm UAV}$ of the GPUs at the UAV, the number of UAVs $n$, and the frame rate $r$.\vspace{-0.33cm}}
    \label{fig:HAP_offloading}
    \vskip -0.2cm
\end{figure*}
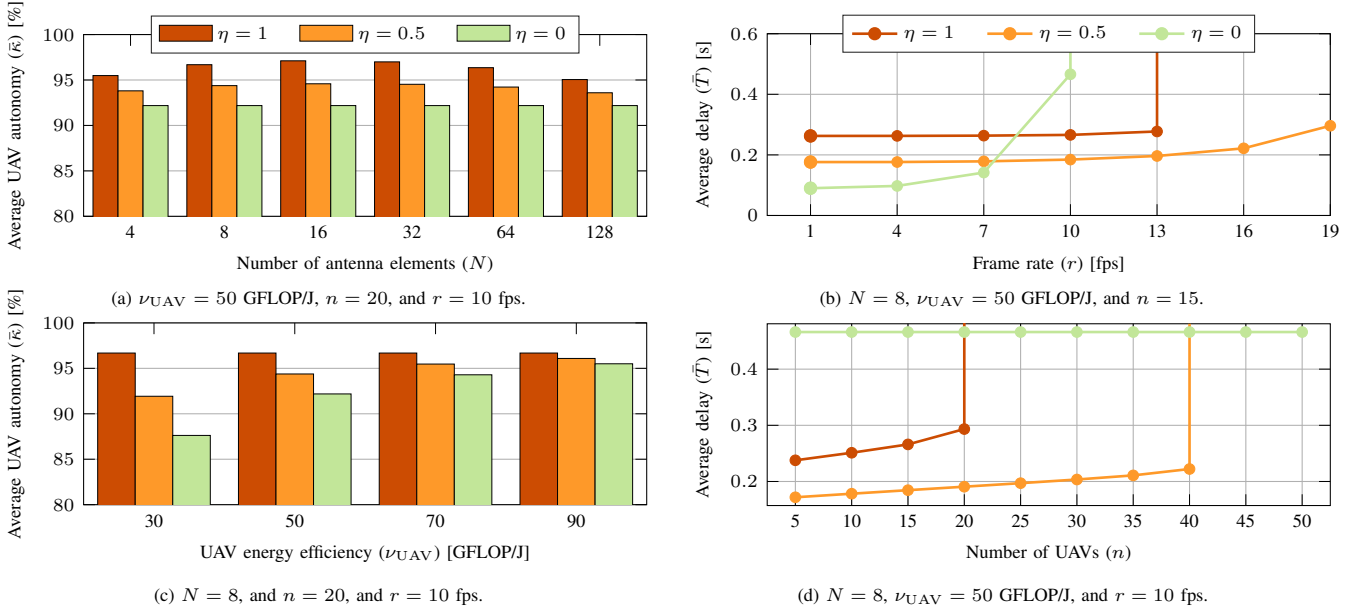

\section{Performance Evaluation}
\label{sec:performance_evaluation}
In~\cref{sec:simulation_setup} we introduce our simulation setup, in~\cref{sec:hap_edge,sec:leo_edge} we present our results for HAP- and LEO-assisted edge computing, respectively, and in~\cref{sec:capacity} we evaluate the energy consumption of the edge~server.

\subsection{Simulation Setup and Parameters}
\label{sec:simulation_setup}
Our main simulation parameters are summarized in~\cref{tab:params1}.

\emph{Communication parameters.} Communication is in the \gls{mmwave} spectrum, specifically in the Ka-bands at 30 (20) GHz in uplink (downlink)~\cite{38811}.
The total bandwidth is $B_{\rm T}=400$~MHz, and the \glspl{uav} operate in orthogonal frequency bands, so the actual bandwidth available to each \gls{uav} is $B=B_{\rm T}/n$.
According to~\cite{38811}, LEO satellites communicate using a \gls{car} antenna, whereas the \gls{hap} is equipped with a \gls{upa} antenna with $N=64$  elements. UAVs also operate with a UPA antenna, and we vary the value of $N$ from 4 to 128.

\emph{Energy parameters.}
The battery capacity of a \gls{uav} is $E^{\rm UAV}_b=130$~Wh, equivalent to that of a DJI Matrice 100~\cite{Woz21Selec}.
Regarding the hovering, from the model in~\cite{Abeywickrama18} we have that $m=3$~kg, $\xi=0.3$~m, and $\psi=1$~kg/m$^3$, so the power consumption in \cref{eq:power_hovering} is around 200 W.
The \gls{hap} and \gls{leo} satellite are powered by batteries with a total capacity of $E^{\rm HAP}_b=8$~kWh and $E^{\rm LEO}_b=6$~kWh, respectively.
As the HAP and the LEO satellite operate above the stratosphere, they can also harvest energy from solar panels of $S_{\rm HAP}=100$~m$^2$ and $S_{\rm LEO}=30$~m$^2$, respectively.

Our scenario consists of a set of $n$ \glspl{uav} that capture images of size $s_{\rm UL}=3$ Mb~\cite{testolina2023selma} at a frame rate $r$, while the processed output, in the form of bounding boxes, is of size $s_{\rm UL}=0.1$~Mb. We assume that all the frames require the same computational load $C_l=90$~GFLOP~\cite{Saeed23OnBoard}. 
UAVs operate with low-cost GPUs (e.g., a Jetson TX2 GPU\footnote{According to the datasheet of a Jetson TX2 GPU~\cite{Bid18Comparing}, the maximum power consumption for data processing is 15~W, the computational capacity is around 1000~GFLOP/s, and the energy efficiency is 41~GFLOP/J.}), so we set the energy efficiency $\nu_{\rm UAV}$ in the range $[30,90]$ GFLOP/J, while the computational capacity is $C_{\rm UAV}=1000$~GFLOP/s. 
On the contrary, the \gls{hap} and the \gls{leo} satellite can support high-performance \glspl{gpu} with a higher energy efficiency of $\nu_{\rm HAP}=\nu_{\rm LEO}=200$~GFLOP/J, and a computational capacity of $C_{\rm HAP}=C_{\rm LEO}=20000$~GFLOP/s.

As far as data offloading is concerned (see~\cref{sub:e-offloading}), using the parameters in \cite{Pizzo17Optimal, Skrimponis20Power}, the total power consumption to transmit and receive data in case of a UPA antenna with $N$ elements is $P^{\rm UPA}_{\rm TX}=P_t+168N+178.5$~mW (with $P_t=30$~dBm) and $P^{\rm UPA}_{\rm RX}=69N+266.8$~mW, respectively, while in case of a \gls{car} antenna we have $P^{\rm CAR}_{\rm TX}=316$~mW and $P^{\rm CAR}_{\rm RX}=305.8$~mW, respectively.



\emph{Performance metrics.}
We evaluate: 
\begin{itemize}
	\item the average delay~$\bar{T}$, measured as in Eq.~\eqref{eq:delay};
	\item the average UAV autonomy $\Bar{\kappa}$, measured as the ratio between the energy consumption for hovering, i.e., $E^{\rm UAV}_{\rm M}$ in~\cref{eq:hovering}, and the overall energy consumption of the UAV, i.e., $E_{\rm UAV}$ in~\cref{eq:e_uav}. As $\Bar{\kappa}$ approaches $1$, the energy consumption is dominated by the sole energy for hovering, which promotes better autonomy;
	\item the energy consumption at the edge server, i.e., $E_{\rm HAP}$ and $E_{\rm LEO}$ in~\cref{eq:e_edge}. 
\end{itemize}

Simulation results are given as a function of the offloading factor $\eta$. We consider: (i) local processing with $\eta = 0$, where the processing is done onboard the \glspl{uav}; (ii) edge computing with $\eta=0.5$, where, on average, half of the frames are processed locally onboard the \glspl{uav} and the rest is offloaded to the edge server; and (iii) edge computing with $\eta = 1$, where all the frames are offloaded to the edge server. Moreover, we investigate the impact of the frame rate $r$, the number of antenna elements $N$ at the UAV, the energy efficiency $\nu_{\rm UAV}$ of the GPUs at the UAV, the elevation angle $\alpha$ of the LEO satellite, the total flying time $t_f$, and the number of UAVs~$n$. Notably, to  maintain the stability of the D/M/1 queues (see~\cref{sub:queue}), the load factor $\rho=\lambda/\mu$ must be less than~1. This sets a limit on the maximum number of \glspl{uav} and the frame rate that the network can handle.
For example, \cref{fig:local_processing_stability} illustrates that the maximum frame rate to maintain stability in case of local processing (i.e., $\eta=0$) is 10~fps, while it is 20~fps if $\eta=0.5$ given that the burden of data processing is partially delegated to the edge server. 
Moreover, in~\cref{fig:50_offloading_stability,fig:100_offloading_stability} we report the load factor $\rho$ vs. $n$ and $r$, which can be used to determine the range of values where stability is guaranteed, i.e., $\rho<1$.
For instance, when $\eta = 0.5$ the network can support up to $20$ UAVs with $r=20$~fps. In turn, when $\eta = 1$, the edge server becomes more congested, and can support only up to  $10$ UAVs with $r=20$~fps.

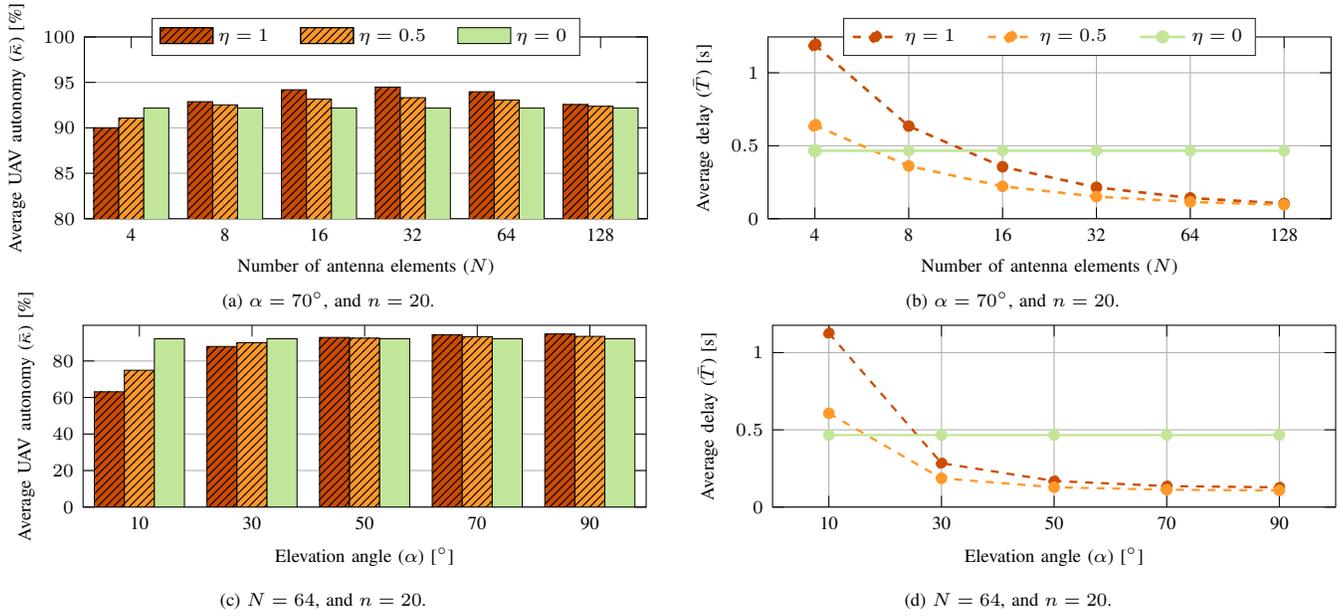
\begin{figure*}[t!]
    \subfloat[][$\alpha=70^\circ$, and $n=20$.\vspace{-0.5cm}]
	{
        \vspace{-0.33cm}
\begin{tikzpicture}
\pgfplotsset{every tick label/.append style={font=\scriptsize}}

\pgfplotsset{compat=1.11,
	/pgfplots/ybar legend/.style={
		/pgfplots/legend image code/.code={%
			\draw[##1,/tikz/.cd,yshift=-0.25em]
			(0cm,0cm) rectangle (20pt,0.6em);},
	},
}

\definecolor{darkgray176}{RGB}{176,176,176}
\definecolor{darkslategray66}{RGB}{66,66,66}
\definecolor{lightgray204}{RGB}{204,204,204}
\definecolor{color1}{RGB}{204,76,2}
\definecolor{color2}{RGB}{254,153,41}
\definecolor{color3}{RGB}{194,230,153}

\begin{axis}[
width = \textwidth/2,
height = 4cm,
legend cell align={left},
legend style={legend cell align=left,
              align=center,
              draw=white!15!black,
              at={(0.5, 1)},
              anchor=center,
              /tikz/every even column/.append style={column sep=1em}},
legend columns=3,
tick pos=both,
unbounded coords=jump,
x grid style={darkgray176},
xlabel={Number of antenna elements (\(\displaystyle N\))},
xmin=-0.5, xmax=5.5,
xtick style={color=black},
xtick={0,1,2,3,4,5},
xticklabels={4,8,16,32,64,128},
y grid style={darkgray176},
ylabel={Average UAV autonomy ($\Bar{\kappa}$) [\%]},
ymajorgrids,
ymin=80, ymax=100,
ytick style={color=black}
]
\draw[draw=black,fill=color1,line width=0.08pt,postaction={pattern=north east lines}, pattern color=black] (axis cs:-0.4,0) rectangle (axis cs:-0.133333333333333,90.0062687405757);
\addlegendimage{ybar,ybar legend,draw=black,fill=color1,line width=0.08pt,postaction={pattern=north east lines}, pattern color=black}
\addlegendentry{$\eta=1$}

\draw[draw=black,fill=color1,line width=0.08pt,postaction={pattern=north east lines}, pattern color=black] (axis cs:0.6,0) rectangle (axis cs:0.866666666666667,92.8774107345834);
\draw[draw=black,fill=color1,line width=0.08pt,postaction={pattern=north east lines}, pattern color=black] (axis cs:1.6,0) rectangle (axis cs:1.86666666666667,94.1865488740954);
\draw[draw=black,fill=color1,line width=0.08pt,postaction={pattern=north east lines}, pattern color=black] (axis cs:2.6,0) rectangle (axis cs:2.86666666666667,94.4853051228489);
\draw[draw=black,fill=color1,line width=0.08pt,postaction={pattern=north east lines}, pattern color=black] (axis cs:3.6,0) rectangle (axis cs:3.86666666666667,93.9704706504702);
\draw[draw=black,fill=color1,line width=0.08pt,postaction={pattern=north east lines}, pattern color=black] (axis cs:4.6,0) rectangle (axis cs:4.86666666666667,92.5700616355004);
\draw[draw=black,fill=color2,line width=0.08pt,postaction={pattern=north east lines}, pattern color=black] (axis cs:-0.133333333333333,0) rectangle (axis cs:0.133333333333333,91.082367704066);
\addlegendimage{ybar,ybar legend,draw=black,fill=color2,line width=0.08pt,postaction={pattern=north east lines}}
\addlegendentry{$\eta=0.5$}

\draw[draw=black,fill=color2,line width=0.08pt,postaction={pattern=north east lines}, pattern color=black] (axis cs:0.866666666666667,0) rectangle (axis cs:1.13333333333333,92.5296628630348);
\draw[draw=black,fill=color2,line width=0.08pt,postaction={pattern=north east lines}, pattern color=black] (axis cs:1.86666666666667,0) rectangle (axis cs:2.13333333333333,93.1747759215266);
\draw[draw=black,fill=color2,line width=0.08pt,postaction={pattern=north east lines}, pattern color=black] (axis cs:2.86666666666667,0) rectangle (axis cs:3.13333333333333,93.320728013171);
\draw[draw=black,fill=color2,line width=0.08pt,postaction={pattern=north east lines}, pattern color=black] (axis cs:3.86666666666667,0) rectangle (axis cs:4.13333333333333,93.0689227771399);
\draw[draw=black,fill=color2,line width=0.08pt,postaction={pattern=north east lines}, pattern color=black] (axis cs:4.86666666666667,0) rectangle (axis cs:5.13333333333333,92.3768831882761);
\draw[draw=black,fill=color3,line width=0.08pt] (axis cs:0.133333333333333,0) rectangle (axis cs:0.4,92.1845093251185);
\addlegendimage{ybar,ybar legend,draw=black,fill=color3,line width=0.08pt}
\addlegendentry{$\eta=0$}

\draw[draw=black,fill=color3,line width=0.08pt] (axis cs:1.13333333333333,0) rectangle (axis cs:1.4,92.1845093251185);
\draw[draw=black,fill=color3,line width=0.08pt] (axis cs:2.13333333333333,0) rectangle (axis cs:2.4,92.1845093251185);
\draw[draw=black,fill=color3,line width=0.08pt] (axis cs:3.13333333333333,0) rectangle (axis cs:3.4,92.1845093251185);
\draw[draw=black,fill=color3,line width=0.08pt] (axis cs:4.13333333333333,0) rectangle (axis cs:4.4,92.1845093251185);
\draw[draw=black,fill=color3,line width=0.08pt] (axis cs:5.13333333333333,0) rectangle (axis cs:5.4,92.1845093251185);
\addplot [line width=0.864pt, darkslategray66, forget plot]
table {%
-0.266666666666667 nan
-0.266666666666667 nan
};
\addplot [line width=0.864pt, darkslategray66, forget plot]
table {%
0.733333333333333 nan
0.733333333333333 nan
};
\addplot [line width=0.864pt, darkslategray66, forget plot]
table {%
1.73333333333333 nan
1.73333333333333 nan
};
\addplot [line width=0.864pt, darkslategray66, forget plot]
table {%
2.73333333333333 nan
2.73333333333333 nan
};
\addplot [line width=0.864pt, darkslategray66, forget plot]
table {%
3.73333333333333 nan
3.73333333333333 nan
};
\addplot [line width=0.864pt, darkslategray66, forget plot]
table {%
4.73333333333333 nan
4.73333333333333 nan
};
\addplot [line width=0.864pt, darkslategray66, forget plot]
table {%
0 nan
0 nan
};
\addplot [line width=0.864pt, darkslategray66, forget plot]
table {%
1 nan
1 nan
};
\addplot [line width=0.864pt, darkslategray66, forget plot]
table {%
2 nan
2 nan
};
\addplot [line width=0.864pt, darkslategray66, forget plot]
table {%
3 nan
3 nan
};
\addplot [line width=0.864pt, darkslategray66, forget plot]
table {%
4 nan
4 nan
};
\addplot [line width=0.864pt, darkslategray66, forget plot]
table {%
5 nan
5 nan
};
\addplot [line width=0.864pt, darkslategray66, forget plot]
table {%
0.266666666666667 nan
0.266666666666667 nan
};
\addplot [line width=0.864pt, darkslategray66, forget plot]
table {%
1.26666666666667 nan
1.26666666666667 nan
};
\addplot [line width=0.864pt, darkslategray66, forget plot]
table {%
2.26666666666667 nan
2.26666666666667 nan
};
\addplot [line width=0.864pt, darkslategray66, forget plot]
table {%
3.26666666666667 nan
3.26666666666667 nan
};
\addplot [line width=0.864pt, darkslategray66, forget plot]
table {%
4.26666666666667 nan
4.26666666666667 nan
};
\addplot [line width=0.864pt, darkslategray66, forget plot]
table {%
5.26666666666667 nan
5.26666666666667 nan
};
\end{axis}

\end{tikzpicture}
        \label{fig:LEO_autonomy_N}
	}
 \hfill
     \subfloat[][$\alpha=70^\circ$, and $n=20$.\vspace{-0.5cm}]
	{
        \vspace{-0.33cm}
\begin{tikzpicture}

\definecolor{darkgray176}{RGB}{176,176,176}
\definecolor{lightgray204}{RGB}{204,204,204}
\definecolor{color1}{RGB}{204,76,2}
\definecolor{color2}{RGB}{254,153,41}
\definecolor{color3}{RGB}{194,230,153}

\begin{axis}[
width = \textwidth/2,
height = 4cm,
legend cell align={left},
legend style={legend cell align=left,
              align=center,
              draw=white!15!black,
              at={(0.5, 1)},
              anchor=center,
              /tikz/every even column/.append style={column sep=1em}},
legend columns=3,
tick pos=both,
unbounded coords=jump,
x grid style={darkgray176},
xlabel={Number of antenna elements (\(\displaystyle N\))},
xmajorgrids,
xmin=-0.5, xmax=5.5,
xtick style={color=black},
xtick={0,1,2,3,4,5},
xticklabels={4,8,16,32,64,128},
y grid style={darkgray176},
ylabel={Average delay ($\Bar{T}$) [s]},
ymajorgrids,
ymin=0., ymax=1.24560397538071,
ytick style={color=black}
]

\addplot [line width=1pt, dashed, color1, mark=*]
table {%
0 1.19085681149854
0 1.19085681149854
};
\addlegendentry{$\eta=1$}

\addplot [line width=1pt, dashed, color2, mark=*]
table {%
0 0.63948717674055
0 0.63948717674055
};
\addlegendentry{$\eta=0.5$}

\addplot [line width=1pt, color3, mark=*]
table {%
0 0.466079347993467
0 0.466079347993467
};
\addlegendentry{$\eta=0$}

\addplot [line width=1pt, color1, dashed, forget plot]
table {%
0 1.19085681149854
1 0.634881137593461
2 0.355927386389358
3 0.214937362360912
4 0.142381555304209
5 0.103709525727639
};
\addplot [line width=1pt, color1, forget plot]
table {%
0 nan
0 nan
};
\addplot [line width=1pt, color1, forget plot]
table {%
1 nan
1 nan
};
\addplot [line width=1pt, color1, forget plot]
table {%
2 nan
2 nan
};
\addplot [line width=1pt, color1, forget plot]
table {%
3 nan
3 nan
};
\addplot [line width=1pt, color1, forget plot]
table {%
4 nan
4 nan
};
\addplot [line width=1pt, color1, forget plot]
table {%
5 nan
5 nan
};
\addplot [draw=color1, fill=color1, mark=*, only marks]
table{%
x  y
0 1.19085681149854
1 0.634881137593461
2 0.355927386389358
3 0.214937362360912
4 0.142381555304209
5 0.103709525727639
};
\addplot [line width=1pt, color2, dashed, forget plot]
table {%
0 0.63948717674055
1 0.361499339788011
2 0.22202246418596
3 0.151527452171737
4 0.115249548643385
5 0.0959135338551
};
\addplot [line width=1pt, color2, forget plot]
table {%
0 nan
0 nan
};
\addplot [line width=1pt, color2, forget plot]
table {%
1 nan
1 nan
};
\addplot [line width=1pt, color2, forget plot]
table {%
2 nan
2 nan
};
\addplot [line width=1pt, color2, forget plot]
table {%
3 nan
3 nan
};
\addplot [line width=1pt, color2, forget plot]
table {%
4 nan
4 nan
};
\addplot [line width=1pt, color2, forget plot]
table {%
5 nan
5 nan
};
\addplot [draw=color2, fill=color2, mark=*, only marks]
table{%
x  y
0 0.63948717674055
1 0.361499339788011
2 0.22202246418596
3 0.151527452171737
4 0.115249548643385
5 0.0959135338551
};
\addplot [line width=1pt, color3, forget plot]
table {%
0 0.466079347993467
1 0.466079347993467
2 0.466079347993467
3 0.466079347993467
4 0.466079347993467
5 0.466079347993467
};
\addplot [line width=1pt, color3, forget plot]
table {%
0 nan
0 nan
};
\addplot [line width=1pt, color3, forget plot]
table {%
1 nan
1 nan
};
\addplot [line width=1pt, color3, forget plot]
table {%
2 nan
2 nan
};
\addplot [line width=1pt, color3, forget plot]
table {%
3 nan
3 nan
};
\addplot [line width=1pt, color3, forget plot]
table {%
4 nan
4 nan
};
\addplot [line width=1pt, color3, forget plot]
table {%
5 nan
5 nan
};
\addplot [draw=color3, fill=color3, mark=*, only marks]
table{%
x  y
0 0.466079347993467
1 0.466079347993467
2 0.466079347993467
3 0.466079347993467
4 0.466079347993467
5 0.466079347993467
};
\end{axis}

\end{tikzpicture}
        \label{fig:LEO_delay_N}
	}
 \vskip 0.1cm
    \subfloat[][$N=64$, and $n=20$.]
	{
		\label{fig:LEO_elev_energy}
\begin{tikzpicture}
\pgfplotsset{every tick label/.append style={font=\scriptsize}}

\pgfplotsset{compat=1.11,
	/pgfplots/ybar legend/.style={
		/pgfplots/legend image code/.code={%
			\draw[##1,/tikz/.cd,yshift=-0.25em]
			(0cm,0cm) rectangle (20pt,0.6em);},
	},
}

\definecolor{darkgray176}{RGB}{176,176,176}
\definecolor{darkslategray66}{RGB}{66,66,66}
\definecolor{lightgray204}{RGB}{204,204,204}
\definecolor{color1}{RGB}{204,76,2}
\definecolor{color2}{RGB}{254,153,41}
\definecolor{color3}{RGB}{194,230,153}

\begin{axis}[
width = \textwidth/2,
height = 4cm,
tick pos=both,
unbounded coords=jump,
x grid style={darkgray176},
xlabel={Elevation angle ($\alpha$) [$^\circ$]},
xmin=-0.5, xmax=4.5,
xtick style={color=black},
xtick={0,1,2,3,4},
xticklabels={10,30,50,70,90},
y grid style={darkgray176},
ylabel={Average UAV autonomy ($\Bar{\kappa}$) [\%]},
ymajorgrids,
ymin=0, ymax=99.5431569714772,
ytick style={color=black}
]
\draw[draw=black,fill=color1,line width=0.08pt,postaction={pattern=north east lines}, pattern color=black] (axis cs:-0.4,0) rectangle (axis cs:-0.133333333333333,63.1135512925187);

\draw[draw=black,fill=color1,line width=0.08pt,postaction={pattern=north east lines}, pattern color=black] (axis cs:0.6,0) rectangle (axis cs:0.866666666666667,87.9470109568931);
\draw[draw=black,fill=color1,line width=0.08pt,postaction={pattern=north east lines}, pattern color=black] (axis cs:1.6,0) rectangle (axis cs:1.86666666666667,92.8979613417948);
\draw[draw=black,fill=color1,line width=0.08pt,postaction={pattern=north east lines}, pattern color=black] (axis cs:2.6,0) rectangle (axis cs:2.86666666666667,94.4229561153592);
\draw[draw=black,fill=color1,line width=0.08pt,postaction={pattern=north east lines}, pattern color=black] (axis cs:3.6,0) rectangle (axis cs:3.86666666666667,94.8030066395021);
\draw[draw=black,fill=color2,line width=0.08pt,postaction={pattern=north east lines}, pattern color=black] (axis cs:-0.133333333333333,0) rectangle (axis cs:0.133333333333333,74.9280671571474);

\draw[draw=black,fill=color2,line width=0.08pt,postaction={pattern=north east lines}, pattern color=black] (axis cs:0.866666666666667,0) rectangle (axis cs:1.13333333333333,90.015917691465);
\draw[draw=black,fill=color2,line width=0.08pt,postaction={pattern=north east lines}, pattern color=black] (axis cs:1.86666666666667,0) rectangle (axis cs:2.13333333333333,92.5398602335396);
\draw[draw=black,fill=color2,line width=0.08pt,postaction={pattern=north east lines}, pattern color=black] (axis cs:2.86666666666667,0) rectangle (axis cs:3.13333333333333,93.2903070943644);
\draw[draw=black,fill=color2,line width=0.08pt,postaction={pattern=north east lines}, pattern color=black] (axis cs:3.86666666666667,0) rectangle (axis cs:4.13333333333333,93.4754237952656);
\draw[draw=black,fill=color3,line width=0.08pt] (axis cs:0.133333333333333,0) rectangle (axis cs:0.4,92.1845093251185);

\draw[draw=black,fill=color3,line width=0.08pt] (axis cs:1.13333333333333,0) rectangle (axis cs:1.4,92.1845093251185);
\draw[draw=black,fill=color3,line width=0.08pt] (axis cs:2.13333333333333,0) rectangle (axis cs:2.4,92.1845093251185);
\draw[draw=black,fill=color3,line width=0.08pt] (axis cs:3.13333333333333,0) rectangle (axis cs:3.4,92.1845093251185);
\draw[draw=black,fill=color3,line width=0.08pt] (axis cs:4.13333333333333,0) rectangle (axis cs:4.4,92.1845093251185);
\addplot [line width=0.864pt, darkslategray66, forget plot]
table {%
-0.266666666666667 nan
-0.266666666666667 nan
};
\addplot [line width=0.864pt, darkslategray66, forget plot]
table {%
0.733333333333333 nan
0.733333333333333 nan
};
\addplot [line width=0.864pt, darkslategray66, forget plot]
table {%
1.73333333333333 nan
1.73333333333333 nan
};
\addplot [line width=0.864pt, darkslategray66, forget plot]
table {%
2.73333333333333 nan
2.73333333333333 nan
};
\addplot [line width=0.864pt, darkslategray66, forget plot]
table {%
3.73333333333333 nan
3.73333333333333 nan
};
\addplot [line width=0.864pt, darkslategray66, forget plot]
table {%
0 nan
0 nan
};
\addplot [line width=0.864pt, darkslategray66, forget plot]
table {%
1 nan
1 nan
};
\addplot [line width=0.864pt, darkslategray66, forget plot]
table {%
2 nan
2 nan
};
\addplot [line width=0.864pt, darkslategray66, forget plot]
table {%
3 nan
3 nan
};
\addplot [line width=0.864pt, darkslategray66, forget plot]
table {%
4 nan
4 nan
};
\addplot [line width=0.864pt, darkslategray66, forget plot]
table {%
0.266666666666667 nan
0.266666666666667 nan
};
\addplot [line width=0.864pt, darkslategray66, forget plot]
table {%
1.26666666666667 nan
1.26666666666667 nan
};
\addplot [line width=0.864pt, darkslategray66, forget plot]
table {%
2.26666666666667 nan
2.26666666666667 nan
};
\addplot [line width=0.864pt, darkslategray66, forget plot]
table {%
3.26666666666667 nan
3.26666666666667 nan
};
\addplot [line width=0.864pt, darkslategray66, forget plot]
table {%
4.26666666666667 nan
4.26666666666667 nan
};
\end{axis}

\end{tikzpicture}
	}
 \hfill
    \subfloat[][$N=64$, and $n=20$.]
	{
		\label{fig:LEO_elev_time}
\begin{tikzpicture}

\definecolor{darkgray176}{RGB}{176,176,176}
\definecolor{lightgray204}{RGB}{204,204,204}
\definecolor{color1}{RGB}{204,76,2}
\definecolor{color2}{RGB}{254,153,41}
\definecolor{color3}{RGB}{194,230,153}
\begin{axis}[
width = \textwidth/2,
height = 4cm,
tick pos=both,
unbounded coords=jump,
x grid style={darkgray176},
xlabel={Elevation angle ($\alpha$) [$^\circ$]},
xmajorgrids,
xmin=-0.5, xmax=4.5,
xtick style={color=black},
xtick={0,1,2,3,4},
xticklabels={10,30,50,70,90},
y grid style={darkgray176},
ylabel={Average delay ($\Bar{T}$) [s]},
ymajorgrids,
ymin=0, ymax=1.17699001878657,
ytick style={color=black}
]
\addplot [line width=0.864pt, color1, dashed, forget plot]
table {%
0 1.12608018330149
1 0.284182957371066
2 0.168996656578053
3 0.13577868997944
4 0.127649405217595
};
\addplot [line width=0.864pt, color1, forget plot]
table {%
0 nan
0 nan
};
\addplot [line width=0.864pt, color1, forget plot]
table {%
1 nan
1 nan
};
\addplot [line width=0.864pt, color1, forget plot]
table {%
2 nan
2 nan
};
\addplot [line width=0.864pt, color1, forget plot]
table {%
3 nan
3 nan
};
\addplot [line width=0.864pt, color1, forget plot]
table {%
4 nan
4 nan
};
\addplot [draw=color1, fill=color1, dashed, mark=*, only marks]
table{%
x  y
0 1.12608018330149
1 0.284182957371066
2 0.168996656578053
3 0.13577868997944
4 0.127649405217595
};
\addplot [line width=0.864pt, color2, dashed, forget plot]
table {%
0 0.607098862642028
1 0.186150249676814
2 0.128557099280307
3 0.111948115981001
4 0.107883473600078
};
\addplot [line width=0.864pt, color2, forget plot]
table {%
0 nan
0 nan
};
\addplot [line width=0.864pt, color2, forget plot]
table {%
1 nan
1 nan
};
\addplot [line width=0.864pt, color2, forget plot]
table {%
2 nan
2 nan
};
\addplot [line width=0.864pt, color2, forget plot]
table {%
3 nan
3 nan
};
\addplot [line width=0.864pt, color2, forget plot]
table {%
4 nan
4 nan
};
\addplot [draw=color2, fill=color2, mark=*, only marks]
table{%
x  y
0 0.607098862642028
1 0.186150249676814
2 0.128557099280307
3 0.111948115981001
4 0.107883473600078
};
\addplot [line width=0.864pt, color3, forget plot]
table {%
0 0.466079347993467
1 0.466079347993467
2 0.466079347993467
3 0.466079347993467
4 0.466079347993467
};
\addplot [line width=0.864pt, color3, forget plot]
table {%
0 nan
0 nan
};
\addplot [line width=0.864pt, color3, forget plot]
table {%
1 nan
1 nan
};
\addplot [line width=0.864pt, color3, forget plot]
table {%
2 nan
2 nan
};
\addplot [line width=0.864pt, color3, forget plot]
table {%
3 nan
3 nan
};
\addplot [line width=0.864pt, color3, forget plot]
table {%
4 nan
4 nan
};
\addplot [draw=color3, fill=color3, mark=*, only marks]
table{%
x  y
0 0.466079347993467
1 0.466079347993467
2 0.466079347993467
3 0.466079347993467
4 0.466079347993467
};
\end{axis}

\end{tikzpicture}
	}
    \caption{LEO-assisted edge computing. Average UAV autonomy $\bar{\kappa}$ (left) and average delay $\bar{T}$ (right) vs. $\eta$, as a function of the elevation angle $\alpha$ of the LEO satellite, and the number of antenna elements $N$ of the UAV.\vspace{-0.33cm}}
    \label{fig:LEO_offloading}
    \vskip -0.2cm
\end{figure*}

\subsection{HAP-Assisted Edge Computing}
\label{sec:hap_edge}
In this section we evaluate the performance of HAP-assisted edge computing vs. local processing.
As expected, \cref{fig:HAP_offloading} shows that data offloading can generally improve the UAV autonomy, but at the expense of a higher delay.
Notably, the impact of $N$ is non-linear, as illustrated in~\cref{fig:HAP_N}.
At first, the UAV autonomy increases as $N$ increases. In fact, the antenna gain produced by beamforming grows with $N$, which permits to improve the quality of the channel and transmit data faster, thereby reducing the energy consumption for data offloading as per~\cref{eq:e_do}.
Then, as $N>16$, the power consumption of the circuitry, which is proportional to $N$ as per \cref{eq10}, keeps increasing but with limited channel improvements, which leads to performance degradation, e.g., for $\eta=1$, the autonomy $\bar{\kappa}$ decreases from 97\% to 93\% as $N$ grows from 16 to 128.

In~\cref{fig:HAP_r} we plot the average delay as a function of the frame rate~$r$. We observe that local processing, which does not involve additional delays for data transmission to the edge server, is the most desirable option as long as $r<7$~fps: the average delay is between 150 and 200~ms.
Then, for $r>10$~fps, the system becomes unstable.
In this case, (partial) data offloading can improve the queuing delay by delegating some processing tasks to the HAP server, despite the communication overhead for uploading data frames. 
In particular, the delay is less than 200 ms in most configurations for $\eta=0.5$, and around 250 ms for  $\eta=1$.

Notice that the energy consumption for local processing depends on the energy efficiency of the GPUs installed at the UAVs.
In~\cref{fig:HAP_eff} we see that using more powerful and efficient (though more expensive) GPUs can reduce the energy consumption for data processing and increase the UAV autonomy by up to $10\%$. Notably, for $\nu_{\rm UAV}=90$~GFLOP/J, $\bar{\kappa}$ is as high as 95\%.

In~\cref{fig:HAP_nUAVs} we see that local processing requires around $450$~ms for each video frame when $N=8$ and $r=10$~fps, regardless of the value of $n$.
On the contrary, the average delay for edge computing via the HAP grows with $n$ due to the more frequent offloading requests to the edge server, and the resulting more populated queues as the number of UAVs increases. In particular, the system becomes unstable as $n \geq 20$ for $\eta=1$, and as $n \geq 40$ for $\eta=0.5$, which is consistent with the results in~\cref{fig:stability}.

\subsection{LEO-Assisted Edge Computing}
\label{sec:leo_edge}
In this section we evaluate the performance of LEO-assisted edge computing vs. local processing.
In general, the delay for data offloading is significantly higher than HAP-assisted edge computing given the longer communication distance to the LEO satellite and the resulting higher path loss.
For example, in \cref{fig:LEO_offloading} we see that the delay for $\eta=1$ can be even higher than 1 s, which is not compatible with the requirements of most wireless applications~\cite{giordani2020toward}.
In this scenario, local processing becomes the most convenient choice in terms of both energy consumption (or UAV autonomy) and delay, especially when $N<4$.
As $N$ increases, the beamforming gain also increases, and the channel quality between the UAV and the LEO satellite improves accordingly. Specifically, the average delay for edge computing for $\eta=1$ decreases from 1120 to only 100 ms when $N$ grows from 4 to 128 elements, with almost no degradation in terms of UAV autonomy, despite the higher power consumption of the circuitry as the number of antennas increases. LEO-assisted edge computing outperforms local processing in terms of both energy consumption (\cref{fig:LEO_autonomy_N}) and delay (\cref{fig:LEO_delay_N}) as $N>16$ (8) for $\eta=1$ (0.5).

Notice that the offloading performance is affected by the elevation angle $\alpha$ between the UAVs and the LEO satellite, as depicted in \cref{fig:LEO_elev_energy,fig:LEO_elev_time}. In particular, as $\alpha$ decreases the path loss increases due to the resulting longer distance between the two endpoints, leading to a higher transmission delay.
For example, for $\alpha=10^\circ$ the UAV autonomy for edge computing ($\eta=1$) is only 60\%, vs. more than 80\% for local processing.
Edge computing becomes more convenient than local processing when $\alpha\geq30^\circ$ in terms of delay, and $\geq50^\circ$ in terms of both UAV autonomy and~delay.

\begin{figure}[t!]
\hspace{0.5 cm}
    \begin{subfigure}[b]{\linewidth}
		\centering
		\setlength\fwidth{\columnwidth}
%
%

\definecolor{black25}{RGB}{25,25,25}
\definecolor{mediumaquamarine102194165}{RGB}{102,194,165}
\definecolor{limegreen3122331}{RGB}{31,223,31}
\definecolor{silver}{RGB}{192,192,192}
\definecolor{lightgreen178223138}{RGB}{178,223,138}
\definecolor{lightsteelblue173203219}{RGB}{173,203,219}
\definecolor{steelblue31120180}{RGB}{31,120,180}

\definecolor{color3}{RGB}{223,101,176}
\definecolor{color1}{RGB}{241,238,246}
\definecolor{color4}{RGB}{206,18,86}
\definecolor{color2}{RGB}{215,181,216}

\begin{tikzpicture}
\pgfplotsset{every tick label/.append style={font=\scriptsize}}

\pgfplotsset{compat=1.11,
	/pgfplots/ybar legend/.style={
		/pgfplots/legend image code/.code={%
			\draw[##1,/tikz/.cd,yshift=-0.25em]
			(0cm,0cm) rectangle (20pt,0.6em);},
	},
}

\begin{axis}[%
width=0,
height=0,
at={(0,0)},
scale only axis,
xmin=0,
xmax=0,
xtick={},
ymin=0,
ymax=0,
ytick={},
axis background/.style={fill=white},
legend style={legend cell align=left,
              align=center,
              draw=white!15!black,
              at={(0, 0)},
              anchor=center,
              /tikz/every even column/.append style={column sep=1em}},
legend columns=3,
]
\addplot[ybar,ybar legend,draw=black,fill=color1,line width=0.08pt]
table[row sep=crcr]{%
	0	0\\
};
\addlegendentry{5 UAVs}

\addplot[ybar legend,ybar,draw=black,fill=color2,line width=0.08pt]
  table[row sep=crcr]{%
	0	0\\
};
\addlegendentry{10 UAVs}

\addplot[ybar legend,ybar,draw=black,fill=color3,line width=0.08pt]
  table[row sep=crcr]{%
	0	0\\
};
\addlegendentry{15 UAVs}

\addplot[ybar legend,ybar,draw=black,fill=color4,line width=0.08pt]
  table[row sep=crcr]{%
	0	0\\
};
\addlegendentry{20 UAVs}

\addplot[ybar legend,ybar,draw=black,fill=white,line width=0.08pt]
  table[row sep=crcr]{%
	0	0\\
};
\addlegendentry{HAP-assisted}

\addplot[ybar legend,ybar,draw=black,fill=white, postaction={pattern=north east lines},line width=0.08pt]
  table[row sep=crcr]{%
	0	0\\
};
\addlegendentry{LEO-assisted}

\end{axis}
\end{tikzpicture}%
	\vspace{-0.33cm}
	\end{subfigure}
   \subfloat
	{
\begin{tikzpicture}

\definecolor{color3}{RGB}{223,101,176}
\definecolor{darkslategray38}{RGB}{38,38,38}
\definecolor{darkslategray66}{RGB}{66,66,66}
\definecolor{lightgray204}{RGB}{204,204,204}
\definecolor{color1}{RGB}{241,238,246}
\definecolor{color4}{RGB}{206,18,86}
\definecolor{color2}{RGB}{215,181,216}

\begin{axis}[
width = \textwidth/2.1,
height = 4cm,
axis line style={lightgray204},
legend cell align={left},
legend style={
  fill opacity=0.8,
  draw opacity=1,
  text opacity=1,
  at={(0.03,0.97)},
  anchor=north west,
  draw=lightgray204
},
tick align=outside,
unbounded coords=jump,
x grid style={lightgray204},
xlabel=\textcolor{darkslategray38}{Total flying time ($t_f$) [min]},
xmajorticks=true,
xmin=-0.5, xmax=5.5,
xtick style={color=darkslategray38},
xtick={0.1,1.1,2.1,3.1,4.1,5.1},
xticklabels={10,20,30,40,50,60},
y grid style={lightgray204},
ylabel=\textcolor{darkslategray38}{Energy consumption ($E$) [Wh]},
ymajorticks=true,
ymajorgrids,
ymin=0, ymax=250,
ytick style={color=darkslategray38}
]
\draw[draw=black,fill=color1,line width=0.08pt] (axis cs:-0.4,0) rectangle (axis cs:-0.2,8.42873414641807);

\draw[draw=black,fill=color1,line width=0.08pt] (axis cs:0.6,0) rectangle (axis cs:0.8,16.8574682928361);
\draw[draw=black,fill=color1,line width=0.08pt] (axis cs:1.6,0) rectangle (axis cs:1.8,25.2862024392542);
\draw[draw=black,fill=color1,line width=0.08pt] (axis cs:2.6,0) rectangle (axis cs:2.8,33.7149365856723);
\draw[draw=black,fill=color1,line width=0.08pt] (axis cs:3.6,0) rectangle (axis cs:3.8,42.1436707320904);
\draw[draw=black,fill=color1,line width=0.08pt] (axis cs:4.6,0) rectangle (axis cs:4.8,50.5724048785084);
\draw[draw=black,fill=color2,line width=0.08pt] (axis cs:-0.2,0) rectangle (axis cs:0,17.8422509758107);

\draw[draw=black,fill=color2,line width=0.08pt] (axis cs:0.8,0) rectangle (axis cs:1,35.6845019516215);
\draw[draw=black,fill=color2,line width=0.08pt] (axis cs:1.8,0) rectangle (axis cs:2,53.5267529274323);
\draw[draw=black,fill=color2,line width=0.08pt] (axis cs:2.8,0) rectangle (axis cs:3,71.369003903243);
\draw[draw=black,fill=color2,line width=0.08pt] (axis cs:3.8,0) rectangle (axis cs:4,89.2112548790538);
\draw[draw=black,fill=color2,line width=0.08pt] (axis cs:4.8,0) rectangle (axis cs:5,107.053505854865);
\draw[draw=black,fill=color3,line width=0.08pt] (axis cs:0,0) rectangle (axis cs:0.2,28.1607416738737);

\draw[draw=black,fill=color3,line width=0.08pt] (axis cs:1,0) rectangle (axis cs:1.2,56.3214833477474);
\draw[draw=black,fill=color3,line width=0.08pt] (axis cs:2,0) rectangle (axis cs:2.2,84.4822250216212);
\draw[draw=black,fill=color3,line width=0.08pt] (axis cs:3,0) rectangle (axis cs:3.2,112.642966695495);
\draw[draw=black,fill=color3,line width=0.08pt] (axis cs:4,0) rectangle (axis cs:4.2,140.803708369369);
\draw[draw=black,fill=color3,line width=0.08pt] (axis cs:5,0) rectangle (axis cs:5.2,168.964450043242);
\draw[draw=black,fill=color4,line width=0.08pt] (axis cs:0.2,0) rectangle (axis cs:0.4,39.3277948604938);

\draw[draw=black,fill=color4,line width=0.08pt] (axis cs:1.2,0) rectangle (axis cs:1.4,78.6555897209877);
\draw[draw=black,fill=color4,line width=0.08pt] (axis cs:2.2,0) rectangle (axis cs:2.4,117.983384581482);
\draw[draw=black,fill=color4,line width=0.08pt] (axis cs:3.2,0) rectangle (axis cs:3.4,157.311179441975);
\draw[draw=black,fill=color4,line width=0.08pt] (axis cs:4.2,0) rectangle (axis cs:4.4,196.638974302469);
\draw[draw=black,fill=color4,line width=0.08pt] (axis cs:5.2,0) rectangle (axis cs:5.4,235.966769162963);


\draw[draw=black,fill=color1,line width=0.08pt, postaction={pattern=north east lines}, pattern color=black] (axis cs:-0.3,0) rectangle (axis cs:-0.1,4.08821522746575);
\draw[draw=black,fill=color1,line width=0.08pt, postaction={pattern=north east lines}, pattern color=black] (axis cs:0.7,0) rectangle (axis cs:0.9,8.17643045493151);
\draw[draw=black,fill=color1,line width=0.08pt, postaction={pattern=north east lines}, pattern color=black] (axis cs:1.7,0) rectangle (axis cs:1.9,12.2646456823973);
\draw[draw=black,fill=color1,line width=0.08pt, postaction={pattern=north east lines}, pattern color=black] (axis cs:2.7,0) rectangle (axis cs:2.9,16.352860909863);
\draw[draw=black,fill=color1,line width=0.08pt, postaction={pattern=north east lines}, pattern color=black] (axis cs:3.7,0) rectangle (axis cs:3.9,20.4410761373288);
\draw[draw=black,fill=color1,line width=0.08pt, postaction={pattern=north east lines}, pattern color=black] (axis cs:4.7,0) rectangle (axis cs:4.9,24.5292913647945);

\draw[draw=black,fill=color2,line width=0.08pt, postaction={pattern=north east lines}, pattern color=black] (axis cs:-0.1,0) rectangle (axis cs:0.1,8.2387990896174);
\draw[draw=black,fill=color2,line width=0.08pt, postaction={pattern=north east lines}, pattern color=black] (axis cs:0.9,0) rectangle (axis cs:1.1,16.4775981792348);
\draw[draw=black,fill=color2,line width=0.08pt, postaction={pattern=north east lines}, pattern color=black] (axis cs:1.9,0) rectangle (axis cs:2.1,24.7163972688522);
\draw[draw=black,fill=color2,line width=0.08pt, postaction={pattern=north east lines}, pattern color=black] (axis cs:2.9,0) rectangle (axis cs:3.1,32.9551963584696);
\draw[draw=black,fill=color2,line width=0.08pt, postaction={pattern=north east lines}, pattern color=black] (axis cs:3.9,0) rectangle (axis cs:4.1,41.193995448087);
\draw[draw=black,fill=color2,line width=0.08pt, postaction={pattern=north east lines}, pattern color=black] (axis cs:4.9,0) rectangle (axis cs:5.1,49.4327945377044);

\draw[draw=black,fill=color3,line width=0.08pt, postaction={pattern=north east lines}, pattern color=black] (axis cs:0.1,0) rectangle (axis cs:0.3,12.44728595157);
\draw[draw=black,fill=color3,line width=0.08pt, postaction={pattern=north east lines}, pattern color=black] (axis cs:1.1,0) rectangle (axis cs:1.3,24.89457190314);
\draw[draw=black,fill=color3,line width=0.08pt, postaction={pattern=north east lines}, pattern color=black] (axis cs:2.1,0) rectangle (axis cs:2.3,37.3418578547099);
\draw[draw=black,fill=color3,line width=0.08pt, postaction={pattern=north east lines}, pattern color=black] (axis cs:3.1,0) rectangle (axis cs:3.3,49.7891438062799);
\draw[draw=black,fill=color3,line width=0.08pt, postaction={pattern=north east lines}, pattern color=black] (axis cs:4.1,0) rectangle (axis cs:4.3,62.2364297578499);
\draw[draw=black,fill=color3,line width=0.08pt, postaction={pattern=north east lines}, pattern color=black] (axis cs:5.1,0) rectangle (axis cs:5.3,74.6837157094199);

\draw[draw=black,fill=color4,line width=0.08pt, postaction={pattern=north east lines}, pattern color=black] (axis cs:0.3,0) rectangle (axis cs:0.5,16.7103653369693);
\draw[draw=black,fill=color4,line width=0.08pt, postaction={pattern=north east lines}, pattern color=black] (axis cs:1.3,0) rectangle (axis cs:1.5,33.4207306739386);
\draw[draw=black,fill=color4,line width=0.08pt, postaction={pattern=north east lines}, pattern color=black] (axis cs:2.3,0) rectangle (axis cs:2.5,50.131096010908);
\draw[draw=black,fill=color4,line width=0.08pt, postaction={pattern=north east lines}, pattern color=black] (axis cs:3.3,0) rectangle (axis cs:3.5,66.8414613478773);
\draw[draw=black,fill=color4,line width=0.08pt, postaction={pattern=north east lines}, pattern color=black] (axis cs:4.3,0) rectangle (axis cs:4.5,83.5518266848466);
\draw[draw=black,fill=color4,line width=0.08pt, postaction={pattern=north east lines}, pattern color=black] (axis cs:5.3,0) rectangle (axis cs:5.5,100.262192021816);

\addplot [line width=0.864pt, darkslategray66, forget plot]
table {%
-0.3 nan
-0.3 nan
};
\addplot [line width=0.864pt, darkslategray66, forget plot]
table {%
0.7 nan
0.7 nan
};
\addplot [line width=0.864pt, darkslategray66, forget plot]
table {%
1.7 nan
1.7 nan
};
\addplot [line width=0.864pt, darkslategray66, forget plot]
table {%
2.7 nan
2.7 nan
};
\addplot [line width=0.864pt, darkslategray66, forget plot]
table {%
3.7 nan
3.7 nan
};
\addplot [line width=0.864pt, darkslategray66, forget plot]
table {%
4.7 nan
4.7 nan
};
\addplot [line width=0.864pt, darkslategray66, forget plot]
table {%
-0.1 nan
-0.1 nan
};
\addplot [line width=0.864pt, darkslategray66, forget plot]
table {%
0.9 nan
0.9 nan
};
\addplot [line width=0.864pt, darkslategray66, forget plot]
table {%
1.9 nan
1.9 nan
};
\addplot [line width=0.864pt, darkslategray66, forget plot]
table {%
2.9 nan
2.9 nan
};
\addplot [line width=0.864pt, darkslategray66, forget plot]
table {%
3.9 nan
3.9 nan
};
\addplot [line width=0.864pt, darkslategray66, forget plot]
table {%
4.9 nan
4.9 nan
};
\addplot [line width=0.864pt, darkslategray66, forget plot]
table {%
0.1 nan
0.1 nan
};
\addplot [line width=0.864pt, darkslategray66, forget plot]
table {%
1.1 nan
1.1 nan
};
\addplot [line width=0.864pt, darkslategray66, forget plot]
table {%
2.1 nan
2.1 nan
};
\addplot [line width=0.864pt, darkslategray66, forget plot]
table {%
3.1 nan
3.1 nan
};
\addplot [line width=0.864pt, darkslategray66, forget plot]
table {%
4.1 nan
4.1 nan
};
\addplot [line width=0.864pt, darkslategray66, forget plot]
table {%
5.1 nan
5.1 nan
};
\addplot [line width=0.864pt, darkslategray66, forget plot]
table {%
0.3 nan
0.3 nan
};
\addplot [line width=0.864pt, darkslategray66, forget plot]
table {%
1.3 nan
1.3 nan
};
\addplot [line width=0.864pt, darkslategray66, forget plot]
table {%
2.3 nan
2.3 nan
};
\addplot [line width=0.864pt, darkslategray66, forget plot]
table {%
3.3 nan
3.3 nan
};
\addplot [line width=0.864pt, darkslategray66, forget plot]
table {%
4.3 nan
4.3 nan
};
\addplot [line width=0.864pt, darkslategray66, forget plot]
table {%
5.3 nan
5.3 nan
};
\end{axis}

\end{tikzpicture}
	}
    \caption{Energy consumption $E_{\rm HAP}$ (plain bar) and $E_{\rm LEO}$ (striped bar) of the edge server for HAP- and LEO-assisted edge computing, respectively, vs. the total flying time $t_f$, as a function of the number of UAVs $n$.\vspace{-0.5cm}}
    \label{fig:avg_capacity}
\end{figure}
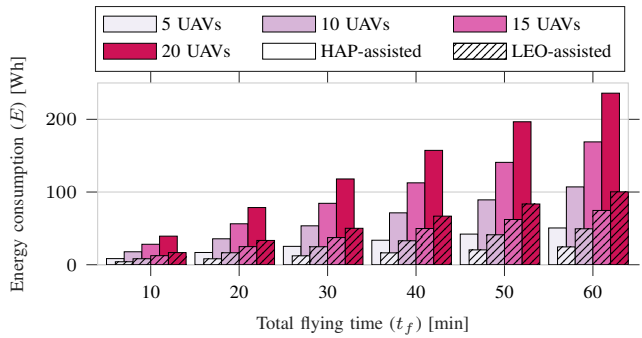

\subsection{Energy Consumption at the Edge Server}
\label{sec:capacity}
In order to evaluate the impact of data offloading, in~\cref{fig:avg_capacity} we show the energy consumption $E_i$ at the edge server $i\in\{\text{HAP, LEO}\}$ as a function of $n$.
We see that the energy consumption at the LEO satellite is lower than at the HAP since the former uses \gls{car} antennas instead of \gls{upa} antennas, which reduces the energy consumption for data transmission and reception (by up to $60\%$ when $n=20$).
As expected, the energy consumption grows over time, and as the number of UAVs increases.
For example, for $n=5$ the LEO satellite (HAP) consumes only 24 (50) Wh after 1~hour of service, while for $n=20$ the energy consumption is more than 4 times higher, given that the edge server has to process more incoming requests. This may saturate the available capacity, which motivates our research towards the optimal offloading factor and configuration for edge computing.

\section{Conclusions and Future Works}
	\label{sec:conclusions_and_future_works}
 In this work, we explored the feasibility of \gls{hap}- and \gls{leo}-assisted edge computing as a solution to process UAV data.
 We evaluated the energy consumption and the delay for data processing considering onboard processing vs. processing at the \gls{hap} or the \gls{leo} satellite. 
We showed that (partial) \gls{hap}-assisted edge computing can improve both autonomy and delay compared to onboard processing in many configurations, and especially when the number of UAVs is lower than 40, and the frame rate is higher than 7 fps. 
In turn, \gls{leo}-assisted edge computing works well only using directional antennas at the \glspl{uav}, and 
with an elevation angle of at least~$50^\circ$.

In our future research, we will formalize an optimization problem with both energy consumption and delay constraints, to identify the optimal offloading factor for edge computing.

\section*{Acknowledgment}
This work was partially supported by the European Union under the Italian National Recovery and Resilience Plan (NRRP) of NextGenerationEU, 
partnership on ``Telecommunications of the Future'' (PE0000001 - program ``RESTART'').

\bibliographystyle{IEEEtran}
\bibliography{bibliography.bib}

\end{document}